\pdfoutput=1
\documentclass[a4paper,11pt]{article}

\usepackage{lineno}
\usepackage{jheppub}
\usepackage[utf8]{inputenc}
\usepackage{amsmath}
\usepackage{amsthm}
\usepackage{amssymb}
\usepackage{enumitem}   
\usepackage[english]{babel}
\usepackage{url}
\usepackage{mathtools}
\usepackage{bbold}
\usepackage{slashed}
\usepackage{multirow}
\usepackage{lipsum}
\usepackage{xcolor}
\usepackage{float}
\usepackage{revsymb}
\usepackage{braket}
\usepackage{siunitx}
\usepackage{tensor}
\usepackage{siunitx}
\usepackage{xfrac}

\DeclareSIUnit\parsec{pc}

\allowdisplaybreaks

\newcommand{\bm}[1]{\boldsymbol{#1}}

\newcommand{\MP}{M_{\rm P}}

\newcommand{\GEF}[2]{\mathcal{F}_\mathcal{{#1}}^{({#2})}}
\newcommand{\dGEF}[2]{\dot{\mathcal{F}}_\mathcal{{#1}}^{({#2})}}
\newcommand{\D}{\mathrm{d}}
\newcommand{\p}{\partial}

\newcommand{\rot}{\mathrm{rot}}

\newcommand{\diver}{\mathrm{div}}

\definecolor{dgreen}{rgb}{0,0.6,0.0}

\begin{document}


\preprint{MS-TP-25-23}

\title{\Large Gravitational waves from axion inflation in the gradient expansion formalism. Part I. Pure axion inflation}

\author[a]{Richard~von~Eckardstein,}
\author[a,b]{Kai~Schmitz,}
\author[a,c]{and Oleksandr~Sobol}

\affiliation[a]{Institute for Theoretical Physics, University of M\"unster,\\
Wilhelm-Klemm-Stra{\ss}e 9, 48149 M\"{u}nster, Germany}

\affiliation[b]{Kavli IPMU (WPI), UTIAS, The University of Tokyo,\\
5-1-5 Kashiwanoha, Kashiwa, Chiba 277-8583, Japan}

\affiliation[c]{Physics Faculty, Taras Shevchenko National University of Kyiv,\\
64/13, Volodymyrska Street, 01601 Kyiv, Ukraine}

\emailAdd{richard.voneckardstein@uni-muenster.de}
\emailAdd{kai.schmitz@uni-muenster.de}
\emailAdd{oleksandr.sobol@uni-muenster.de}


\abstract{Axion inflation is a well-motivated model of cosmic inflation with a rich phenomenology. The abundant production of gauge fields during axion inflation notably sources a stochastic gravitational-wave (GW) background signal, which nourishes the hope that future GW searches might have a chance to probe the model. In this paper, we scrutinize GW production during axion inflation in the gradient expansion formalism (GEF), a powerful numerical technique that captures the nonlinear dynamics of the system in the limit of vanishing axion gradients. We focus on {single-field} axion inflation coupled to a pure Abelian gauge sector, i.e., pure axion inflation (PAI), and perform the first-ever {detailed} parameter scan of GW production in the Abelian PAI model close to the onset of strong backreaction. {We approximate the axion potential around its minimum by a quadratic mass term and study the tensor modes that exit the Hubble horizon as the axion rolls down this potential.} Remarkably enough, we find that GW signals within the reach of future GW interferometers can only be realized in parameter regions that also lead to strong backreaction and that are in conflict with the upper limit on $\Delta N_{\rm eff}$, i.e., the allowed energy density of dark radiation. This observation defines a clear target for future lattice studies of axion inflation that may confirm or improve the predictions of our GEF benchmark.}


\maketitle


\section{Introduction}
\label{sec: introduction}


The direct detection of gravitational waves (GWs)~\cite{abbott_2016}, a hundred years after their theoretical prediction~\cite{Einstein_1916, Einstein_1918}, has paved the way towards the next major milestone in fundamental physics: the detection of a stochastic GW background (SGWB). Indeed, recent results from pulsar timing arrays (PTAs) already show varying evidence in favor of its existence~\cite{agazie_nanograv_2023,epta+inpta_2023,reardon_PPTA_2023,xu_CPTA_2023,miles_meerkat_2024}. The detection of an SGWB promises an exciting opportunity to probe and constrain the physics of the early Universe before Big Bang nucleosynthesis (BBN) and the emission of the cosmic microwave background (CMB), enabling one to probe energy scales far beyond those of the Standard Model (SM) of particle physics (for a review, see Ref.~\cite{caprini_2018}).

A particularly intriguing prospect for searches of an SGWB is that the detection of a signal could shed light on the physics of cosmic inflation~\cite{starobinsky_1980,guth_1981,linde_1982, albrecht_1982, starobinsky_1982,linde_1983}. Not only is inflation the leading theory to resolve the horizon, flatness, and monopole problems of the Hot Big Bang model, it is also a source of primordial perturbations in the Universe's energy density and spacetime metric~\cite{Mukhanov_1981,Mukhanov_1982, hawking_1982, guth_1982, bardeen_1983}, which can explain the temperature anisotropies of the CMB. At the same time, tensor perturbations of the metric would source an SGWB~\cite{Grishchuk_1974,Starobinsky_1979,rubakov_1982,fabbri_1983, abbott_1984}, which could be as well detected by future CMB missions~\cite{abazajian_cmb-s4_2022}, but also by PTAs~\cite{agazie_nanograv_RPL_2025, afzal_nanograv_NP_2023} and other GW observatories~\cite{guzzetti_2016}. Furthermore, nonminimal particle dynamics during inflation and effects at second order in cosmological perturbation theory can result in extra contributions to the SGWB, e.g., GWs sourced by gauge fields (i.e., gauge-field-induced GWs or GFIGWs for short) or scalar-induced GWs (SIGWs), which could lead to a detectable SGWB at high frequencies (for reviews, see Refs.~\cite{guzzetti_2016, domenech_scalar_2021}). The detection of these signals may help to constrain inflationary physics at frequency scales far above the CMB scale.

A model of cosmic inflation with an exceptionally rich phenomenology is axion inflation~\cite{garretson_1992,anber_2006, Anber_2010}. This extension of the standard inflationary slow-roll paradigm builds upon the idea of natural inflation~\cite{freese_1990} by assuming inflation to be driven by an axion-like pseudoscalar inflaton field, or simply axion field for short, whose potential is protected against radiative corrections by an approximate shift symmetry. An inflaton field $\phi$ of this kind naturally couples to gauge fields via operators of the form $\phi\,F_{\mu\nu}\tilde{F}^{\mu\nu}$, where the field strength tensor $F_{\mu\nu}$ and its dual $\tilde{F}^{\mu\nu}$ belong to an Abelian gauge field $A_\mu$ in the simplest types of models. In this paper, we will consider axion inflation of exactly this type.

The axion--vector coupling leads to the abundant production of helical gauge fields~\cite{garretson_1992} due to the spontaneous parity violation from the rolling axion field, which can have implications for magnetogenesis~\cite{garretson_1992, anber_2006, durrer_2011, bamba_2015, fujita_2015, adshead_2016} and reheating~\cite{barnaby_2012, adshead_2015, Cheng_2016, adshead_2018,cuissa_2019,adshead_2020_A,adshead_2020_B,sharma_2025}. Gauge-field production can also induce significant non-Gaussian scalar and tensor perturbations~\cite{Anber_2010, Barnaby_2011_A, barnaby_2012, Sorbo_2011, cook_2012, linde_2013, bugaev_2014, ferreira_2014, Cheng_2016, domcke_2016, garcia-bellido_2016, adshead_2020_A, adshead_2020_B, Domcke_2020_Resonant, corba_2024, durrer_2024, greco_2024,sharma_2025}. The former may source SIGWs and primordial black holes (PBHs)~\cite{linde_2013, bugaev_2014, Cheng_2016, Domcke_2020_Resonant, ozsoy_2023, sharma_2025,talebian_2025}, but are also constrained by the CMB~\cite{barnaby_2012, linde_2013, bugaev_2014, ferreira_2014}, while the latter may source substantial amounts of GWs, which could be detected by future interferometer experiments or PTAs~\cite{barnaby_2012, cook_2012, domcke_2016,garcia-bellido_2024}. Meanwhile, the production of GFIGWs could be even too efficient, resulting in an overproduction of gravitational radiation in conflict with CMB and BBN bounds on additional dark radiation beyond the SM photons and neutrinos~\cite{adshead_2020_A, adshead_2020_B, sharma_2025}. 

Importantly, the production of Abelian gauge fields can impact the dynamics of the inflaton field by providing a source of friction in addition to the usual Hubble friction caused by the cosmic expansion~\cite{Anber_2010}. The so-called strong-backreaction regime, where gauge-field friction is important, leads to an instability in the dynamics of the system, resulting in a highly non-linear inflationary evolution~\cite{Cheng_2016, Domcke_2020_Resonant, peloso_2023, caravano_2022, eckardstein_2023, Figueroa_2023, caravano_2023, domcke_2024, Figueroa_2024}. Recently, lattice simulations demonstrated that these effects can lead to sizable axion gradients that additionally alter the evolution of the gauge-field--inflaton system~\cite{Figueroa_2023, Figueroa_2024, sharma_2025}.

The strong gauge fields present during axion inflation could also entail the pair creation of charged scalars and fermions via the Schwinger effect~\cite{domcke_2018, Sobol_2019, Domcke_2020_Fermions, Gorbar_2021,gorbar_2022,fujita_2022, cado_2022, domcke_2023,gorbar_2023, bastero-gil_2024_A, bastero-gil_2024_B, eckardstein_2025, iarygina_2025}.
If the $U(1)$ gauge field is identified with the SM hypercharge field, and the fermions charged under the $U(1)$ gauge symmetry therefore correspond to the fermions of the SM, the dual production of gauge fields and fermions during axion inflation can have important implications for baryogenesis~\cite{anber_2015, adshead_2015,jimenez_2017, cado_2022, domcke_2023}.
The generation of fermionic matter has also been considered as a potential model of inflationary dark matter production~\cite{bastero-gil_2024_A,bastero-gil_2024_B}. To differentiate between models of axion inflation with and without fermion production, we refer to the former as \textit{fermionic axion inflation} (FAI), and to the latter as \textit{pure axion inflation} (PAI). In particular, the model considered in this paper, axion inflation coupled to a pure Abelian gauge sector, may be referred to as Abelian PAI; a companion paper deals with Abelian FAI~\cite{eckardstein_2025a}. 

The goal of this paper is to study the production of GFIGWs during Abelian PAI in the transition regime between weak and strong gauge-field backreaction utilizing the so-called gradient-expansion formalism (GEF)~\cite{Sobol_2019, Gorbar_2021, gorbar_2022, durrer_2023, eckardstein_2023, domcke_2024}, a powerful numerical method that has been used to validate and calibrate numerical lattice simulations of axion inflation in the past~\cite{Figueroa_2023,sharma_2025,Figueroa_2024}. Our GEF analysis of GW production during Abelian PAI in this paper can thus be regarded as a precursor investigation of more detailed lattice studies. In other words, the present work provides a GEF benchmark for future lattice studies highlighting the potentially most interesting regions in parameter space.

In particular, we shall focus on the observable prospects for planned interferometry experiments, such as the Einstein telescope (ET)~\cite{ET_2010} and the Laser Interferometer Space Antenna (LISA)~\cite{LISA_2017, LISA_2019}.
We constrain GW production by accounting for the BBN and CMB constraints on dark relativistic degrees of freedom, $\Delta N_{\mathrm{eff}}$~\cite{planck_2020_VI, yeh_2021, pisanti_2021}, and the non-observation of an SGWB during the third observation run of the ground-based interferometer collaborations LIGO~\cite{LIGO_2010, LIGO_2014} and Virgo~\cite{VIRGO_2014} (HLVO3)~\cite{LIGO_collaboration_2021}. Throughout this article, we will study a class of inflationary models whose scalar potential is quadratic during the last $e$-folds of inflation. As we will demonstrate, it turns out that an observable GW signal within the reach of ET or LISA can only be attained in parameter regions that also result in strong backreaction. This statement is worth being repeated: We find that an observable GW signal only comes at the cost of strong backreaction. Simultaneously, we observe that the same backreaction effects lead to the overproduction of GWs in tension with $\Delta N_{\mathrm{eff}}$ bounds. Together, these two observations lead to the drastic conclusion that\,---\,within the context of our GEF benchmark\,---\,the detection prospects of GFIGWs from PAI are limited. Because of the limitations of our numerical approach, this statement does not amount to a general no-go theorem. After all, in our GEF approach, we model strong-backreaction effects by assuming a homogeneous inflaton evolution, which enables us to perform an extensive parameter scan. A full treatment would require the inclusion of inflaton gradients, which is currently only possible with lattice simulations.
In this sense, our GEF benchmark defines a clear target for future lattice studies that may or may not confirm the incompatibility between an observable GW signal and bounds on $\Delta N_{\mathrm{eff}}$ that we observe in our case.

The rest of the paper is organized as follows: In Secs.~\ref{sec: GFIGW} and \ref{sec: axion inflation}, we will review the computation of the SGWB signal from gauge fields and the PAI model, respectively. In Sec.~\ref{sec: Parameter space}, we will discuss the modeling of the inflationary dynamics, CMB constraints, and detection characteristics for GW interferometers. Sec.~\ref{sec: Constraints on strong BR} presents the results of an extensive parameter space scan leading to the conclusion that strong homogeneous backreaction is in tension with bounds on $\Delta N_{\mathrm{eff}}$. Finally, in Sec.~\ref{sec: conclusions}, we will summarize our findings and provide a brief outlook. For technical details on our numerical methods, see Appendix~\ref{app: numerics}.

\medskip\noindent
\textbf{Notation:} Throughout this paper, we assume the background spacetime to be described by the spatially flat Friedmann--Lema\^{i}tre--Robertson--Walker (FLRW) metric,
\begin{equation}
    ds^2 = \tensor{g}{_\mu_\nu} \D x^\mu \D x^\nu = \D t^2 - a^2(t) \tensor{\delta}{_i_j} \D x^i \D x^j = a^2(\eta)\left(\D \eta^2 -  \tensor{\delta}{_i_j}\D x^i \D x^j \right)\, , 
    \label{eq: FLRW}
\end{equation}
with scale factor $a$, physical time $t$, conformal time $\eta$, Greek indices for four-vectors, Latin indices for Euclidean three-vectors, and metric signature $(+,-,-,-)$. Unless stated otherwise, $f'$ denotes the derivative of $f$ with respect to conformal time, while $\dot{f}$ denotes the derivative with respect to physical time. We define the Levi-Civita symbol in four dimensions as $\tensor{\varepsilon}{^0^1^2^3} = 1$.  Throughout this paper, we work in natural units, $c = \hbar = 1$, and define the reduced Planck mass as $\MP = 1/\sqrt{8 \pi\,G} \simeq \SI{2.435e18}{\giga\eV}$.


\section{Gravitational waves induced by gauge fields}
\label{sec: GFIGW}


We begin our discussion by reviewing the production of GWs induced by gauge fields
~\cite{Barnaby_2011_A, barnaby_2012, Sorbo_2011,garcia-bellido_2024}. 
Consider the action
\begin{equation}
    S = \int \D^4 x\, \sqrt{-g} \left(- \frac{\MP^2}{2}R+\mathcal{L}_M \right) \, ,
    \label{eq: EH action}
\end{equation}
with $R$ the Ricci scalar and $\mathcal{L}_M$ the Lagrangian density describing the matter part of the action. 
We are working under the assumption of a perturbed FLRW spacetime, keeping only the transverse--traceless metric perturbations,
$\tensor*{h}{_i_j^{\mathrm{TT}}}$, up to first order,\footnote{At first order in cosmological perturbation theory, scalar, vector and tensor perturbations are decoupled from each other, so we are free to treat tensor perturbations separately.}
\begin{equation}
    ds^2 = a^2(\eta) \left( \D \eta^2-  \left[\tensor{\delta}{_i_j} + \tensor*{h}{_i_j^{\mathrm{TT}}}\right]\D x^i \D x^j \right) \, .
\end{equation}
Assuming the perturbative ansatz for the metric holds, it implies that the Einstein field equations split into a background and a perturbed equation as
\begin{equation}
    \langle \tensor*{G}{^\mu_\nu} \rangle = \frac{1}{\MP^2}\langle \tensor*{T}{^\mu_\nu} \rangle\,, \qquad
    \left(\tensor*{G}{^\mu_\nu} - \langle \tensor*{G}{^\mu_\nu} \rangle \right) = \frac{1}{\MP^2}\left(\tensor*{T}{^\mu_\nu} - \langle \tensor*{T}{^\mu_\nu} \rangle\right) \, ,
    \label{eq: Einstein}
\end{equation}
where $\tensor*{G}{^\mu_\nu}$ is the Einstein tensor and $\tensor*{T}{^\mu_\nu}$ the energy--momentum tensor obtained by varying the matter part of the action in Eq.~\eqref{eq: EH action} with respect to the metric.
This separation splits the homogeneous part of the field equations from their inhomogeneous, perturbed part,
such that the homogeneous part is given by the Friedmann equations,
\begin{equation}
    \mathcal{H}^2 = \frac{a^2 \bar\rho}{3 \MP^2}\, ,\qquad (2\mathcal{H}' + \mathcal{H}^2) = -\frac{a^2 \bar p}{\MP^2} \, ,
    \label{eq: Friedmann}
\end{equation}
with $\mathcal{H} = a'/a$ the comoving Hubble rate,  and $\bar \rho = \langle \tensor*{T}{^0_0} \rangle$ and $\bar{p} = - \sfrac{1}{3}\,\langle \tensor*{\delta}{^i_j}\tensor*{T}{^j_i} \rangle$ the energy density and pressure of the homogeneous background system.
The expectation value $\langle \cdot \rangle$ prescribes how the matter part of the system should be split into homogeneous and perturbed contributions. For a quantum system, $\langle \cdot \rangle$ is the quantum expectation value, whereas for a classical system, it is the spatial average. In our case, we will deal with a quantum system (i.e., mode functions of a quantized vector field originating from the Bunch--Davies vacuum) that decoheres over the course of inflation, ultimately giving rise to classical electric and magnetic fields on cosmological scales. In the course of this quantum-to-classical transition during inflation, the conceptual interpretation of $\langle \cdot \rangle$ changes accordingly.

The equation of motion (EOM) of the tensor perturbations is given by
\begin{equation}
    \left(\tensor*{h}{_i_j^{\mathrm{TT}}}\right)'' + 2 \mathcal{H} \left(\tensor*{h}{_i_j^{\mathrm{TT}}}\right)' - \Delta \tensor*{h}{_i_j^{\mathrm{TT}}} = \frac{2}{\MP^2} a^2 \tensor*{\sigma}{_i_j^{\mathrm{TT}}} \,,
    \label{eq: hijTTEOM}
\end{equation}
where $\Delta = \p_i \p^i$ is the spatial Laplacian, and $\tensor*{\sigma}{_i_j^{\mathrm{TT}}}$ is the transverse--traceless part of the energy--momentum tensor of the matter fields,
\begin{equation}
    \tensor*{\sigma}{_i_j^{\mathrm{TT}}} = - \tensor{\delta}{_i_k} \left(\tensor*{T}{^k_j} - \langle \tensor*{T}{^k_j} \rangle\right)^{\mathrm{TT}} = -\tensor{\delta}{_i_k}\left(\tensor*{T}{^k_j}\right)^{\mathrm{TT}} \,,
\end{equation}
where we used $\langle \tensor*{T}{^i_j} \rangle^{\mathrm{TT}}=0$, assuming $\langle \cdot \rangle$ projects out the homogeneous background.

Following these general considerations, we are ready to specify the matter Lagrangian $\mathcal{L}_M$, stating that it contains a term describing a single species of Abelian gauge fields,
\begin{equation}
    \mathcal{L}_M \supset -\frac{1}{4} \tensor{F}{_\mu_\nu} \tensor{F}{^\mu^\nu} \, ,
    \label{eq: GF Lagrangian}
\end{equation}
where $\tensor{F}{_\mu_\nu} = \p_\mu A_\nu - \p_\nu A_\mu$ denotes the field-strength tensor belonging to the comoving four-vector potential $A_\mu$. The gauge-field contribution to the anisotropic stress is given by
\begin{equation}
    \tensor*{(\sigma_{\mathrm{EM}})}{_i_j^{\mathrm{TT}}} = - \left(E_i E_j +  B_i B_j \right)^{\mathrm{TT}} \, ,
    \label{eq: GF anisotropic stress}
\end{equation}
as can be easily derived by varying Eq.~$\eqref{eq: GF Lagrangian}$ with respect to the metric. The electric field $\bm{E} = (E_1,E_2, E_3)$ and magnetic field $\bm{B} = (B_1,B_2, B_3)$ are defined as%
\footnote{For simplicity, we shall refer to $\bm{E}$ and $\bm{B}$ as electric and magnetic fields, although they are defined for arbitrary Abelian gauge fields and do not necessarily belong to the ${\rm U}(1)_{\mathrm{EM}}$ gauge field (photon) of the SM.}

\begin{equation}
    \tensor{F}{_0_i} = a^2 E_i, \quad \tensor{F}{_i_j} = - a^2 \tensor{\varepsilon}{_i_j_k} B^k \, .
    \label{eq: E&B definition}
\end{equation}
The appearance of the FLRW scale factor $a$ in these definitions ensures that $\bm{E}$ and $\bm{B}$ correspond to the physical fields measured by a comoving observer.

We are interested in situations where gauge fields are the dominant source of GWs. Thus, we will now proceed by assuming $\tensor*{\sigma}{_i_j^{\mathrm{TT}}} = \tensor*{(\sigma_{\mathrm{EM}})}{_i_j^{\mathrm{TT}}}$, 
such that Eq.~\eqref{eq: hijTTEOM} turns into
\begin{equation}
    \left(\tensor*{h}{_i_j^{\mathrm{TT}}}\right)'' + 2 \mathcal{H} \left(\tensor*{h}{_i_j^{\mathrm{TT}}}\right)' - \Delta \tensor*{h}{_i_j^{\mathrm{TT}}} = -\frac{2}{\MP^2} a^2 \left(E_i E_j +  B_i B_j \right)^{\mathrm{TT}} \, .
\end{equation}
In order to solve this EOM, it is advantageous to fix a gauge. We will ultimately be interested in gauge fields solely produced via their interaction with the homogeneous inflationary background, but from no other source, such as, e.g., classical charges or currents. We are thus free to treat the electric and magnetic fields in the far-field limit and work in radiation gauge, which combines the Coulomb gauge condition, $\diver \bm{A} = 0$, with the temporal (or Weyl) gauge condition, $A_0 = 0$. Furthermore, we treat the gauge field as a quantum operator, writing it in Fourier space as
\begin{equation}
    \hat{\bm{A}}(\eta, \bm{x}) = \int \frac{\D^3 \bm{k}}{(2 \pi)^{3/2}}\sum_{\lambda={\pm1}} 
    \left( \bm{\epsilon}_\lambda(\bm{k}) A_\lambda(\eta, k) \hat{a}^{\vphantom{\dagger}}_\lambda(\bm{k})e^{i \bm{k} \cdot \bm{x}} 
    + \bm{\epsilon}_\lambda^*(\bm{k}) A^*_\lambda(\eta, k) \hat{a}_\lambda^\dagger(\bm{k})e^{-i \bm{k} \cdot \bm{x}} \right).
    \label{eq: GF FT}
\end{equation}
Here, we have chosen a helicity basis, with polarization vectors $\bm{\epsilon}_\lambda(\bm{k})$ defined such that
\begin{equation}
    \bm{k} \cdot \bm{\epsilon}_\lambda(\bm{k}) = 0, \quad i \bm{k} \times \bm{\epsilon}_\lambda(\bm{k}) 
    = \lambda k\, \bm{\epsilon}_\lambda(\bm{k}), \quad \bm{\epsilon}_\lambda(\bm{k})\cdot\bm{\epsilon}_{\lambda'}^*(\bm{k}) = \delta_{\lambda \lambda'}, \quad \bm{\epsilon}^\ast_\lambda(\bm{k})=\bm{\epsilon}_{-\lambda}(\bm{k})=\bm{\epsilon}_{\lambda}(-\bm{k})\, .
\end{equation}
The creation and annihilation operators $\hat{a}^\dagger_\lambda(\bm{k})$ and $\hat{a}^{\vphantom{\dagger}}_\lambda(\bm{k})$ in Eq.~\eqref{eq: GF FT} satisfy the canonical commutation relations for the ladder operators of the quantum harmonic oscillator,
\begin{equation}
    [\hat{a}^{\vphantom{\dagger}}_\lambda(\bm{k}), \hat{a}^\dagger_{\lambda'}(\bm{k'})] = \delta_{\lambda \lambda'} \delta^{(3)}(\bm{k} - \bm{k'}), \quad
     [\hat{a}^{\vphantom{\dagger}}_\lambda(\bm{k}), \hat{a}^{\vphantom{\dagger}}_{\lambda'}(\bm{k'})] = [\hat{a}^\dagger_\lambda(\bm{k}), \hat{a}^\dagger_{\lambda'}(\bm{k'})] = 0 \, .
    \label{eq: canonical commutation relation}
\end{equation}
Upon going to Fourier space, we can write the tensor perturbation as
\begin{equation}
    \tensor*{\hat{h}}{_i_j^{\mathrm{TT}}}(\eta, \bm{x}) 
    = \int \frac{\D^3 \bm{k}}{(2 \pi)^{3/2}} \sum_{\lambda={\pm1}}\left( \epsilon_i^\lambda(\bm{k}) \epsilon_j^\lambda(\bm{k}) \hat{h}_\lambda(\eta, k) e^{i \bm{k} \cdot \bm{x}} + \text{h.c.} \right)\, .
    \label{eq:hijTT}
\end{equation}
Similarly as the vector field, we will treat $\tensor*{h}{^{\mathrm{TT}}_{ij}}$ as a quantum field. The results summarized in the remainder of this section thus apply to quantum systems whose dynamical degrees of freedom correspond to a $U(1)$ vector field and tensor metric perturbations. The model of Abelian PAI discussed in the next section will be an important application of these results; however, much of the remaining discussion in this section can also be readily applied to other cosmological scenarios. 
In Eq.~\eqref{eq:hijTT}, we utilized our freedom to pick a polarization basis for the tensor perturbations, opting for a helicity basis to match the polarization basis of the gauge fields. It is then straightforward to derive the EOM for $\hat{h}_\lambda$,
\begin{equation}
    \hat{h}_\lambda'' + 2 \mathcal{H} \hat{h}_\lambda' + k^2 \hat{h}_\lambda =
     -\frac{2}{\MP^2} a^2 \tensor*{\Pi}{_\lambda^{ij}}(\bm{k})  \int \frac{\D^3 \bm{x}}{(2 \pi)^{3/2}} \left(\hat{E}_i \hat{E}_j +  \hat{B}_i \hat{B}_j \right) e^{-i \bm{k}\cdot\bm{x}} \, ,
    \label{eq: GW EoM, Fourier}
\end{equation}
where $\tensor*{\Pi}{_\lambda^{ij}} = \epsilon^i_{-\lambda}(\bm{k}) \epsilon^j_{-\lambda}(\bm{k})$ is the projector onto the helical polarization basis for $\lambda = \pm 1$. 
We also used that projecting onto the helicity basis is equivalent to projecting onto the transverse--traceless component of a tensor,
thus we can drop the superscript $TT$ for the source term on the right-hand side of Eq.~\eqref{eq: GW EoM, Fourier}.

From here, we can find solutions to Eq.~\eqref{eq: GW EoM, Fourier} by combining its homogeneous and particular solutions, 
$\hat{h}_\lambda = \sfrac{2}{\MP}\left(\hat{u}_\lambda^{\mathrm{vac}} +  \hat{u}_\lambda^{\mathrm{ind}}\right)$. We refer to these two contributions as the vacuum and the induced contribution, respectively.
The rescaled perturbations $\hat{u}_\lambda = \sfrac{\MP}{2}\, \hat{h}_\lambda$ are defined such that the second-order perturbed Einstein--Hilbert action expressed through $u$ contains a canonical kinetic term.
Focusing first on the vacuum contribution, we define
\begin{equation}
    \hat{u}_\lambda^{\mathrm{vac}}(\eta, \bm{k}) = u_\lambda^{0}(\eta, k) \hat{b}^{\vphantom{\dagger}}_\lambda(\bm{k}) + {u_\lambda^{0}}^*(\eta, k) \hat{b}^\dagger_\lambda(-\bm{k})\,,   
\end{equation}
where the creation and annihilation operators $\hat{b}^\dagger_\lambda$ and $\hat{b}^{\vphantom{\dagger}}_\lambda$ are canonically commuting analogously to Eq.~\eqref{eq: canonical commutation relation}, 
and the vacuum modes $u_\lambda^0(\eta, k)$ obey the damped wave equation
\begin{equation}
    \mathcal{D}_k {u_\lambda^0} = {u_\lambda^0}'' + 2 \mathcal{H} {u_\lambda^0}' + k^2 {u_\lambda^0} = 0\, .
    \label{eq: GW vacuum mode EoM}
\end{equation}
The induced contribution $\hat{u}_\lambda^{\mathrm{ind}}$ is then given by
\begin{equation}
    \hat{u}_\lambda^{\mathrm{ind}}(\eta, k) =  -\frac{1}{\MP}\int_{-\infty}^{0} \D \tau \, a^2(\tau) G_k(\eta, \tau) \tensor*{\Pi}{^i^j_\lambda}(\bm{k})
      \int \frac{\D^3 \bm{x}}{(2 \pi)^{3/2}} \left(\hat{E}_i \hat{E}_j +  \hat{B}_i \hat{B}_j \right) e^{-i \bm{k}\cdot\bm{x}}\, ,
\end{equation}
with $G_k(\eta, \tau)$ the retarded Green function for the operator $\mathcal{D}_k$,
\begin{equation}
    G_k(\eta, \tau) 
    = \frac{ \operatorname{Im} \left[ {u_\lambda^0}^*(\eta, k) \, u_\lambda^0(\tau,k) \right]  }
    { \operatorname{Im} \left[ {u_\lambda^0}(\tau, k) \, {{u_\lambda^0}^*}'(\tau,k) \right]   }  \theta (\eta - \tau)\, .
    \label{eq: Green function}
\end{equation}
Continuing from this point, we define the polarized tensor power spectrum through\footnote{Our convention for the polarized tensor power spectrum $\mathcal{P}_{T,\lambda}$
differs from the one used in previous works, e.g., Ref.~\cite{garcia-bellido_2024}.
The two conventions are related by $\mathcal{P}^{\rm this\,  work}_{T,\lambda} = 2\,\mathcal{P}^{\rm other\,  convention}_{T,\lambda}$. 
The unpolarized power spectra are the same in both conventions, 
$\mathcal{P}_{T} = (\sum_\lambda \mathcal{P}^{\rm this\, work}_{T,\lambda})/2 
                    = \sum_\lambda \mathcal{P}^{\rm other\,  convention}_{T,\lambda}$ }
\begin{equation}
    \langle \hat{h}_\lambda(\eta,\bm{k}) \hat{h}_{\lambda'}(\eta,\bm{k'}) \rangle 
    = \delta^{(3)}(\bm{k} + \bm{k'}) \delta_{\lambda\lambda'} \frac{\pi^2}{k^3} \mathcal{P}_{T,\lambda}(k) \,.
    \label{eq: polarised PT}
\end{equation}
Expanding $\hat{h}_\lambda$ in terms of a vacuum and an induced contribution, we can apply Wick's theorem, from which we obtain that the power spectrum splits into two contributions
\begin{equation}
        \mathcal{P}_{T,\lambda}(\eta, k) = \mathcal{P}_{T,\lambda}^{\mathrm{vac}}(\eta, k) + \mathcal{P}_{T,\lambda}^{\mathrm{ind}}(\eta, k)\, ,
\end{equation}
where
\begin{subequations}
    \begin{align}
         \mathcal{P}_{T,\lambda}^{\mathrm{vac}}(\eta, k) &= \frac{4 k^3}{\pi^2 \MP^2} |u_\lambda^0(\eta, k)|^2\, ,\\
         \mathcal{P}_{T,\lambda}^{\mathrm{ind}}(\eta, k) &= \frac{k^3}{2 \pi^2 \MP^4} \int \frac{\D^3 \bm{p}}{(2 \pi)^3} \sum_{\alpha,\beta = \pm1} 
            \left(1 +  \lambda \alpha \frac{\bm{k} \cdot \bm{p}}{k p} \right)^2 \left(1 +  \lambda \beta \frac{k^2 - \bm{k} \cdot \bm{p}}{kq}  \right)^2 \\ 
        &\times \left|\int_{-\infty}^0 \D \tau \frac{G_k(\eta, \tau)}{a^2(\tau)} 
            \left[A'_\alpha(\tau, p)A'_\beta(\tau, q) + \alpha \beta\, p q\, A_\alpha(\tau, p) A_\alpha(\tau, q) \right]
                    \right|^2 \nonumber \, ,
    \end{align}%
    \label{eq: PT}%
\end{subequations}%
with the relative momentum $q = |\bm{k} - \bm{p}|$. Notably, there are no cross-terms between vacuum and induced tensor perturbations. Below, we will be interested in the generation of GFIGWs during a period of slow-roll inflation. Then, the vacuum contribution to the tensor power spectrum is approximately given by
\begin{equation}
    \mathcal{P}_{T, \lambda}^{\mathrm{vac}} =\mathcal{P}_{T}^{\mathrm{vac}} \simeq 2 \left( \frac{H}{\pi \MP} \right)^2 \, ,
    \label{eq: vacuum power spectrum}
\end{equation}
while the induced contribution depends on the mode functions $A_\lambda(\eta, k)$. In fact, once the background dynamics are given, this is the only strong model dependence in Eq.~\eqref{eq: PT}.
From the polarized power spectrum, one obtains the total tensor power spectrum as
\begin{equation}
    \mathcal{P}_{T} = \frac{1}{2} \left(\mathcal{P}_{T,+} + \mathcal{P}_{T,-}\right) = \mathcal{P}_T^{\mathrm{vac}} + \frac{1}{2} \left( \mathcal{P}_{T,+}^{\mathrm{ind}} + \mathcal{P}_{T,-}^{\mathrm{ind}}\right)\, .
\end{equation}
%

The quantity we are ultimately interested in is the GW energy-density power spectrum, i.e., the GW energy density per logarithmic frequency in units of the critical energy density,
\begin{equation}
    \Omega_\mathrm{GW}(f) = \frac{1}{3 H_0^2 \MP^2}\frac{\D \rho_\mathrm{GW} (f)}{\D \ln{f}} \, ,
\end{equation}
where $H_0=h\times \SI{100}{\kilo\metre\per\second\per\mega\parsec}$ is the Hubble rate today. $\Omega_\mathrm{GW}(f)$ can be directly obtained from the tensor power spectrum by accounting for two effects: cosmological redshift of the GW frequency and the evolution of the amplitudes of the individual GW modes through the FLRW spacetime after their source has switched off~\cite{Boyle_2008}. 
The redshift in frequency between the time of emission and today is given by $f = k/(2\pi a_0)$, where $a_0$ is the scale factor today.
To quantify the second effect, consider the evolution of a GW mode with wavenumber $k$. 
Assuming that, following its production, the mode exits the Hubble horizon at a time $\eta_{\mathrm{out}}$, $k \simeq \mathcal{H}(\eta_{\mathrm{out}})$,
this mode will re-enter the horizon at a later time when $k \simeq \mathcal{H}(\eta_{\mathrm{in}})$, with the time of re-entry,  $\eta_{\mathrm{in}}$, depending on the wavenumber $k$. Following the re-entry inside the Hubble horizon, we assume the source of GW production has switched off, so the mode will follow the evolution of a damped harmonic oscillator given by Eq.~\eqref{eq: GW vacuum mode EoM}. The solution of the EOM can be cast in the form of a transfer function that relates the initial mode amplitudes set during inflation to their present-day values~\cite{caprini_2018},
\begin{equation}
    |\mathcal{T}_\mathrm{GW}(f)|^2 \simeq \frac{H_0^2\Omega_r}{8 \pi^2 f^2} \frac{g_*(T_f)}{g_*(T_0)} \left(\frac{g_{*,S}(T_0)}{g_{*,S}(T_f)}\right)^{4/3} \,.
    \label{eq:transfer}
\end{equation}
Here, $\Omega_r$ is the fractional energy density in radiation today, and $g_*(T)$ and $g_{*,S}(T)$ count the effective numbers of relativistic degrees of freedom contributing to the energy density and entropy density of the thermal plasma at temperature $T$, respectively. The transfer function notably obtains its form in Eq.~\eqref{eq:transfer} after invoking entropy conservation in the expanding Universe, $g_{*,S}\,a^3T^3 = \textrm{const}$. Finally, $T_0 \simeq \SI{2.73}{K}$ is the temperature of the CMB photons today,  and $T_f$ is the temperature of the SM plasma when the frequency $f$ re-entered the horizon. Combining the two effects described above, we obtain
\begin{equation}
    \Omega_\mathrm{GW}(f) = \frac{1}{24} \Omega_r \frac{g_*(T_f)}{g_*(T_0)} \left(\frac{g_{*,S}(T_0)}{g_{*,S}(T_f)}\right)^{4/3} \mathcal{P}_T(\eta_{\mathrm{out}}(k_f), k_f) \,, \quad k_f = 2 \pi a_0 f \,.
    \label{eq: GW from PT}
\end{equation}

The maximal frequency reached by GWs produced during inflation corresponds to modes leaving the Hubble horizon at the end of inflation, $k = a_{\mathrm{end}} H_{\mathrm{end}}$,
\begin{align}
    f_{\mathrm{end}} = \frac{H_{\mathrm{end}}}{2 \pi} \left(\frac{a_{\mathrm{end}}}{a_{\mathrm{reh}}}\right) \left( \frac{g_{*,S}(T_0)}{g_{*,S}(T_{\mathrm{reh}})}\right)^{1/3} \frac{T_0}{T_{\mathrm{reh}}} \overset{{\rm inst. \, reh.}}{\simeq} \num{1.83e11} \left(\frac{H_{\mathrm{end}}}{\MP}\right)^{1/2} \SI{}{\hertz} \, .
    \label{eq: f_end}
\end{align}

Here, we used that the evolution of the scale factor below the reheating temperature, $T_{\mathrm{reh}}$, may be inferred from entropy conservation, and that the initial phase of reheating lasts for $\ln ({a_{\mathrm{reh}}/a_{\mathrm{end}}})$ $e$-folds. The last estimate is obtained assuming instantaneous reheating, i.e., $T_{\mathrm{reh}}^{4} = 90 H_{\mathrm{end}}^2 \MP^2/(\pi^2 g_{*}(T_{\mathrm{reh}}))$ and using $g_*(T_{\mathrm{reh}}) = g_{*,S}(T_{\mathrm{reh}}) \simeq 104.4$ for $T_{\mathrm{reh}} \gtrsim \SI{2e2}{\GeV}$~\cite{saikawa_2018, saikawa_2020}.

Up to this point, we managed to carry out the calculation without concerning ourselves with a specific model for gauge-field production. However, at this stage, we need to specify the precise nature and origin of the gauge fields, which leads us to the discussion of axion inflation in the next section.


\section{Pure axion inflation}
\label{sec: axion inflation}


Our model of axion inflation consists of an Abelian gauge field coupled via a Chern--Simons-type interaction to a pseudoscalar inflaton field, $\phi$. This specifies the matter Lagrangian of the action in Eq.~\eqref{eq: EH action} as
\begin{equation}
    \mathcal{L}_M = \frac{1}{2} g^{\mu \nu} \p_{\mu} \phi \,  \p_{\nu} \phi - V(\phi) 
        - \frac{1}{4} \tensor{F}{_\mu_\nu} \tensor{F}{^\mu^\nu} - \frac{1}{4} I(\phi) \tensor{F}{_\mu_\nu} \tensor{\tilde{F}}{^\mu^\nu} ,
    \label{eq: Lagrangian, generic axion inflation}
\end{equation}
where $V(\phi)$ is an inflaton potential, $\tensor{\tilde{F}}{^\mu^\nu} = \tensor{\varepsilon}{^\mu^\nu^\alpha^\beta}\,\tensor{F}{_\alpha_\beta}/(2\sqrt{-g})$ is the dual to the field strength tensor, and
$I(\phi)$ is a generic axial coupling function. 

The energy--momentum tensor for this system takes contributions from its two constituents,
the inflaton field and the gauge fields,
\begin{equation}
    \tensor{T}{_\mu_\nu} = \frac{1}{2}\p_{\mu} \phi \, \p_{\nu} \phi -\tensor{F}{_\mu^\alpha} \tensor{F}{_\alpha_\nu} 
    - g_{\mu \nu} \left(\frac{1}{2} \p_{\alpha} \phi \,  \p^{\alpha} \phi - V(\phi) - \frac{1}{4} \tensor{F}{_\alpha_\beta} \tensor{F}{^\alpha^\beta}\right) \,.
\end{equation}
The background energy density and pressure are then given by
\begin{equation}
        \bar{\rho} = \frac{1}{2} {\dot\varphi}^2 + V(\varphi) + \frac{1}{2} \langle \bm{E}^2 + \bm{B}^2 \rangle, 
        \qquad 
        \bar{p} = \frac{1}{2} {\dot\varphi}^2 - V(\varphi) + \frac{1}{6} \langle \bm{E}^2 + \bm{B}^2 \rangle \,.
        \label{eq: Background energy density and pressure}
\end{equation}
Here, we assumed that spatial gradients in the inflaton field are negligibly small, implying that all expectation values of the inflaton field may be expressed in terms of its time-evolving zero mode, $\varphi(t) = \langle \phi(t, \bm{x})\rangle$. 
By splitting the energy--momentum tensor into a homogeneous contribution describing pressure and energy density at the background level on the one hand as well as perturbations around this background on the other hand, 
we find that the anisotropic stress of the system is given by $\tensor*{\sigma}{_i_j^{\mathrm{TT}}} \simeq \tensor*{(\sigma_{\mathrm{EM}})}{_i_j^{\mathrm{TT}}}$, 
with $\tensor*{(\sigma_{\mathrm{EM}})}{_i_j^{\mathrm{TT}}}$ given in Eq.~\eqref{eq: GF anisotropic stress}. 
Assuming $\nabla \phi \approx 0$ and working at first order in perturbation theory, the relevant sources of GWs in our model exactly correspond to those discussed in the previous section. In principle, axion gradients, and more generally any type of scalar perturbation, also act as a source of GWs. However, the production of SIGWs is an effect at second order in perturbation theory and hence negligible in the limit $\nabla \phi \rightarrow 0$. Consequently, the tensor power spectrum is given entirely by the expression in Eq.~\eqref{eq: PT}.

With this prescription for separating background quantities from perturbations,
the EOMs derived from Eq.~\eqref{eq: Lagrangian, generic axion inflation} lead to the following set of equations to describe the dynamics of the background,
\begin{subequations}
    \begin{equation}
        \Ddot{\varphi} + 3H\dot{\varphi} + V_{,\phi}(\varphi) = \frac{1}{2} I_{,\phi}(\varphi) \braket{\bm{E} \cdot \bm{B} + \bm{B} \cdot \bm{E}}\, ,
        \label{eq: phiEoM}
    \end{equation}
    \begin{equation}
        \diver \bm{E} = 0\, , \qquad \diver \bm{B} = 0\, ,
    \end{equation}
    \begin{equation}
        \dot{\bm{E}} + 2 H \bm{E} - \frac{1}{a}\rot \bm{B} + I_{,\phi}(\varphi)  \dot{\varphi}\bm{B} = 0 \, ,
        \label{eq: EEoM}
    \end{equation}
    \begin{equation}
        \dot{\bm{B}}  + 2 H \bm{B} + \frac{1}{a}\rot \bm{E} = 0 \, .
        \label{eq: BEoM}
    \end{equation}%
    \label{eq: EoMs}%
\end{subequations}%
These equations are closed by supplementing them with the Friedmann equation, Eq.~\eqref{eq: Friedmann} and the total energy density, Eq.~\eqref{eq: Background energy density and pressure}.

The Abelian PAI model  features a remarkably abundant helical gauge-field production, as can be most easily understood from the evolution equation of the gauge-field modes $A_\lambda(\eta, k)$ defined in a helicity basis as in Eq.~\eqref{eq: GF FT},
\begin{equation}
    A_\lambda''(\eta, k)  + \left( k^2 - 2 \lambda k \xi \mathcal{H} \right) A_\lambda(\eta, k) = 0 \,.
    \label{eq: Mode Eq - Pure}
\end{equation}
Evidently, only one of the helicities $\lambda = \pm 1$ is tachyonically amplified depending on the sign of the instability parameter (or gauge-field production parameter)  $\xi = I_{,\phi}(\varphi) \dot{\varphi}/(2H)$.

It is well known that, assuming a perfect de Sitter expansion and an adiabatically slowly varying inflaton velocity, such that $\xi \simeq {\rm const}$, Eq.~\eqref{eq: Mode Eq - Pure} may be integrated exactly.
Using these solutions in Eq.~\eqref{eq: PT}, the tensor power spectrum is~\cite{Barnaby_2011_A, barnaby_2012,Sorbo_2011}
\begin{equation}
    \mathcal{P}_T^{\mathrm{ind}} \:\:\overset{\rm no \, BR}{\simeq}\:\: \num{8.6e-7} \left( \frac{H}{\MP} \right)^4 \frac{\exp (4\pi |\xi|) }{\pi^2 |\xi|^6} \,,
    \label{eq: PT analytical}
\end{equation}
where ``no BR'' indicates that this relation holds when backreaction effects are negligible. This analytical estimate may be compared to the vacuum contribution, Eq.~\eqref{eq: vacuum power spectrum}, which scales quadratically with $H$. 
However, even for low $H$, the induced contribution may be significantly enhanced due to gauge-field production controlled by the parameter $\xi$. Thus, as the inflaton velocity typically increases towards the end of inflation, the induced contribution can start to dominate over the vacuum term towards the end of inflation.

The abundant production of gauge fields may, however, crucially impact the dynamics of the inflaton field $\varphi$.
By sourcing the term $\langle \bm{E} \cdot \bm{B}\rangle$, gauge fields can impart a sizable amount of friction onto the inflaton field~\cite{Anber_2010}, thereby rendering the typical slow-roll attractor solution unstable~\cite{peloso_2023, eckardstein_2023}. This regime of non-linear axion--gauge-field dynamics is known as the strong-backreaction regime. Quantitatively, conditions to detect the onset of this regime may be determined \textit{a posteriori} by studying the impact of the friction term $\langle \bm{E} \cdot \bm{B} \rangle$ over Hubble friction, 
or the contribution of the electromagnetic energy density to the total energy density. These comparisons can be parametrized in terms of the backreaction parameters
\begin{subequations}
    \begin{equation}
        \delta_{\mathrm{KG}} = \frac{|I_{,\phi}(\varphi) \langle \bm{E} \cdot \bm{B} \rangle |}{ |3 H \dot{\varphi}| } \:\:\overset{\rm no \, BR}{\simeq}\:\: \num{4.1e-5} (I_{,\phi}(\varphi) H)^2 \frac{e^{2 \pi |\xi|}}{|\xi|^5} \,, 
    \end{equation}
    \begin{equation}
        \delta_{\mathrm{F}} = \frac{\langle \bm{E}^2 \rangle + \langle \bm{B}^2 \rangle}{ 6 H^2 \MP^2 } \:\:\overset{\rm no \, BR}{\simeq}\:\: \num{4.1e-5} \left(\frac{H}{\MP}\right)^2 \frac{e^{2 \pi |\xi|}}{|\xi|^5} (1 + 1.1 |\xi|^2) \, .
    \end{equation}%
    \label{eq: BR parameters}%
\end{subequations}%
Again, ``no BR'' indicates that these results only apply to the case of weak backreaction.

The strong-backreaction regime induces oscillations in the inflaton velocity $\dot{\varphi}$ due to a retarded response between gauge-field production and backreaction~\cite{Domcke_2020_Resonant}. Since the backreaction drains kinetic energy from the inflaton, this oscillatory stage also extends inflation past its expected duration based purely on slow-roll dynamics. 
The inflationary stage instead is typically terminated when the gauge-field energy density has grown sufficiently to match the energy density in the inflaton potential, $V(\varphi) \sim \rho_{\mathrm{EM}}$, rather than by the kinetic energy of the inflaton field, $V(\varphi) \sim \dot{\varphi}^2$. As a consequence, inflation is immediately followed by a stage of radiation domination.
While the onset of the strong-backreaction regime has been studied analytically~\cite{Domcke_2020_Resonant,peloso_2023}, 
numerical methods are necessary to fully describe the non-linear dynamics of the axion--gauge-field system.
Three different strategies have been applied to achieve this: 
(i) solving the gauge-field dynamics in momentum space~\cite{fujita_2015, Cheng_2016, Notari_2016, Domcke_2020_Resonant,bastero-gil_2022, garcia-bellido_2024}, i.e., integrating Eq.~\eqref{eq: Mode Eq - Pure} in parallel to Eq.~\eqref{eq: phiEoM}; (ii) solving the entire system in Eq.~\eqref{eq: EoMs} using lattice techniques~\cite{adshead_2015, adshead_2016, Figueroa_2018, cuissa_2019, adshead_2018, adshead_2020_A, adshead_2020_B, caravano_2022, Figueroa_2023, caravano_2023, Figueroa_2024, sharma_2025, Jamieson:2025ngu}; 
or (iii) a method known as the gradient-expansion formalism (GEF)~\cite{Sobol_2019, Gorbar_2021, gorbar_2022, durrer_2023, eckardstein_2023, domcke_2024} which is centered around directly solving evolution equations for bilinear expectation values of the form $\langle \bm{E} \cdot \rot^n \bm{E} \rangle$,  $\langle \bm{E} \cdot \rot^n \bm{B} \rangle$ and $\langle \bm{B} \cdot \rot^n \bm{B} \rangle$.

It should be noted that the assumption of neglecting inhomogeneities in the axion field breaks down in the strong-backreaction regime, as demonstrated by recent lattice simulations~\cite{Figueroa_2023, Figueroa_2024, sharma_2025}. Axion gradients are sourced by the gauge field and can drastically alter the dynamics of PAI in the strong-backreaction regime. Specifically, one finds that the oscillations in the inflaton velocity are damped, entailing a more gradual gauge-field production. Nonetheless, the duration of inflation is prolonged until the gauge-field energy density contributes a large fraction of the total energy density. Methods relying on the assumption of a fully homogeneous axion field (GEF and momentum-space methods) fail to accurately reproduce these dynamics of strong backreaction found on the lattice after the first oscillation in the inflaton velocity. In particular, attempts at including axion inhomogeneities in the GEF have only led to minor improvements so far~\cite{domcke_2024}. Still, homogeneous methods are relevant for phenomenological parameter scans due to their larger dynamical range and reduced computational cost. The qualitative agreement between all available methods still remains at a reasonable level, such that GEF benchmarks, as the one worked out in the present paper, are suitable to define targets for future lattice studies. 


\section{Model specifics and observational constraints}
\label{sec: Parameter space}


\subsection{Modeling of the inflationary dynamics}

To study the production of GFIGWs during PAI, we need to specify the axion--vector coupling $I(\phi)$ and the inflaton potential $V(\phi)$. 
Throughout this article, we restrict ourselves to a standard linear axion--vector coupling, $I(\phi) = (\beta/\MP) \phi$, parametrized such that $\beta/\MP = \alpha/f$, where $\alpha$ is the axion--vector coupling constant and $f$ is the axion decay constant. Furthermore, we assume a quadratic potential for the inflaton, $V(\phi) = m^2 \phi^2 / 2$.

Although a quadratic inflation potential is ruled out by CMB measurements \cite{planck_2020_X}, we adhere to this potential for two reasons. First, this potential is the primary benchmark scenario in the literature on strong gauge-field backreaction. Second, it yields representative results for GFIGW spectra for any inflationary potential ending its slow-roll trajectory in a nearly quadratic potential.
This is due to the large separation between frequencies covered by current and planned GW observatories,
$f_\mathrm{GW} \sim \num{e-9} - \SI{e6}{\hertz}$, and the frequency scales constrained by the CMB measurements, $f_{\mathrm{CMB}} \sim \SI{e-18}{\hertz}$.
Our results are relevant for the class of inflationary models with scalar potentials of approximately the form
\begin{equation}
    V(\varphi)  = 
        \left\{
            \begin{array}{lll}
            \frac{1}{2} m^2 \varphi^2 &, & \quad |\varphi| \lesssim \varphi_{\mathrm{thr}} \\
            V_{\mathrm{CMB}}(\varphi) &, & \quad |\varphi| \gtrsim \varphi_{\mathrm{thr}}  \\
            \end{array}
        \right.  \:\:,
        \label{eq: heuristic inflationary potential}
\end{equation}
i.e., having an arbitrary potential shape $V_{\mathrm{CMB}}(\varphi)$ at early times, such that at $\varphi_{\mathrm{CMB}}$ the potential satisfies CMB constraints, and a quadratic potential as the inflaton approaches the end of inflation. As long as the gauge fields only affect the inflaton sufficiently late, long after the moment when $\left|\varphi\right| \sim \varphi_{\mathrm{thr}}$, our results for GW production close to the end of inflation will be insensitive to the choice of $V_{\mathrm{CMB}}(\varphi)$. Let us explain this point in more detail: Around the transition in the shape of the potential, when backreaction effects from gauge-field production are still negligible, the slow-roll evolution of the inflation field will approach the attractor solution of chaotic inflation~\cite{Salopek_1990, liddle_1994}. Then, for an appropriate choice of the threshold value $\varphi_{\rm thr}$, gauge-field production will only become relevant after the system has already spent a few $e$-folds in the attractor solution. Hence, the subsequent inflaton evolution, gauge-field production, and SGWB spectra will have no memory of the model-dependent shape of the potential at large field values. In other words, our results derived at $|\varphi| \lesssim \varphi_{\mathrm{thr}}$ will be identical to those of a pure $V(\phi) = m^2 \phi^2 / 2$ model. The conditions imposed by our construction on $V_{\mathrm{CMB}}(\varphi)$ are minimal: continuity demands $V_{\mathrm{CMB}}(\varphi_{\mathrm{thr}}) = m^2 \varphi_{\mathrm{thr}}^2/2$, and monotonicity imposes $V_{\mathrm{CMB}}(\varphi_{\mathrm{CMB}}) \gtrsim V_{\mathrm{CMB}}(\varphi_{\mathrm{thr}})$.

One may constrain the mass of the inflaton, $m$, by using standard slow-roll results~\cite{dodelsonModernCosmology2021} for the amplitude of scalar perturbations, $A_S$, and the tensor-to-scalar ratio, $r$,
\begin{equation}
    A_S \simeq \left. \frac{V}{24 \pi^2 \MP^4 \varepsilon_V} \right\vert_{\varphi = \varphi_{\mathrm{CMB}}}, \qquad r = \left. 16 \varepsilon_V \right\vert_{\varphi = \varphi_{\mathrm{CMB}}} \,, 
\end{equation}
with the slow-roll parameter $\varepsilon_V = (\MP V_{,\varphi}/V)^2/2$. Using the measured value $A_S = \exp(3.044) \num{e-10}=\num{2.1e-9}$ \cite{planck_2020_VI,galloni_2024}, and the upper bound, $r \lesssim 0.03$, \cite{tristram_2022, galloni_2024}
one arrives at
\begin{equation}
    V(\varphi_{\mathrm{thr}}) \lesssim  V_{\mathrm{CMB}}(\varphi_{\mathrm{CMB}}) \lesssim \num{9.3e-10} \MP^4\,.
    \label{eq: tensor-to-scalar ratio bound}
\end{equation}
A second constraint can be derived by considering the bound on non-Gaussianities in the scalar power spectrum  at CMB scales.
This upper bound may be written as $|\xi_{\mathrm{CMB}}| \lesssim 2.5$~\cite{Barnaby_2011_A,barnaby_2012,Planck:2015zfm,Planck:2019kim}. Expressing $\xi$ in terms of the slow-roll parameter $\varepsilon_V$, we can reformulate these constraints into the condition $\varepsilon_V \lesssim 12.5/\beta^2$. 
When combined with the measured value of $A_S$, this condition can be reformulated into a constraint on the inflaton potential at $\varphi_{\mathrm{thr}}$
\begin{equation}
    V(\varphi_{\mathrm{thr}}) \lesssim  V_{\mathrm{CMB}}(\varphi_{\mathrm{CMB}}) \lesssim \num{6.2e-6} \MP^4/\beta^2 \,.
    \label{eq: non-Gaussianity bound}
\end{equation}

Next, we seek to translate these bounds on $V(\varphi_{\mathrm{thr}})$ into bounds on the inflaton mass $m$. To do so, it is worth remembering the goal of this exercise: we are interested in determining at which point the inflaton potential will have to be quadratic to a good approximation, such that GW spectra that are computed assuming a quadratic potential are indeed insensitive to the potential shape at higher scales. To do so, one can translate $\varphi_{\mathrm{thr}}$ to a characteristic threshold frequency, $f_{\mathrm{thr}}$, which corresponds to the frequency of GWs today that left the horizon when $\varphi = \varphi_{\mathrm{thr}}$.
This way, GWs with frequencies above $f_{\mathrm{thr}}$ are sure to have originated from the inflationary dynamics evaluated in a quadratic potential.
By using standard results for the dynamics of the inflaton field during chaotic inflation, one can approximately relate this threshold frequency to the inflaton potential at the threshold scale,
\begin{equation}
    V(\varphi_{\mathrm{thr}}) \simeq m^2 \MP^2 \left[51.7 - 4.6 \log_{10} \left( \frac{f_{\mathrm{thr}}}{\SI{}{\hertz}} \right) + 2.3 \log_{10} \left( \frac{m}{\MP} \right) - 2 \Delta N_{\mathrm{BR}} \right] \, .
    \label{eq: V threshold}
\end{equation}
To arrive at this equation, we use that any GW frequency $f$ can be related to its corresponding comoving momentum $k$ as $f = k / (2 \pi a_0)$, and that the largest GW frequency, corresponding to GWs produced at the end of inflation, is given in terms of the Hubble rate at the end of inflation (see Eq.~\eqref{eq: f_end}, which, for chaotic inflation, is $H_{\mathrm{end}} \simeq m/\sqrt{6}$. To account for the possibility of strong-backreaction effects prolonging inflation past the expected slow-roll result, we also include an additional red-shift term $\Delta N_{\mathrm{BR}}$. While in Eq.~\eqref{eq: V threshold}, we assume that reheating occurs instantaneously, the effect of reheating would also be an additional source of red-shift, entering into Eq.~\eqref{eq: V threshold} the same way as $\Delta N_{\mathrm{BR}}$.

To see Eq.~\eqref{eq: V threshold} applied, consider two scenarios (first for $\Delta N_{\mathrm{BR}}=0$): In a conservative construction, we may demand that the GW spectrum starting from low frequencies upwards needs to originate from the chaotic-inflation stage of Eq.~\eqref{eq: heuristic inflationary potential}. In this case, the CMB constraints from Eq.~\eqref{eq: tensor-to-scalar ratio bound} imposes a tighter bound on the inflaton mass $m$, as there is less flexibility in modifying the inflaton potential at higher scales to accommodate CMB constraints. For example, choosing the threshold frequency to correspond to energy scales relevant for BBN, $f_{\mathrm{thr}} = f_{_\mathrm{BBN}} \simeq \SI{e-12}{\hertz}$, we find $m \lesssim \num{3.1e-6} \MP$.
In the opposite extreme, we may state that only GW spectra at high frequencies need to originate from chaotic inflation dynamics, picking, e.g., $f_{\mathrm{thr}} = \SI{e-6}{\hertz}$. Then the bound by Eq.~\eqref{eq: tensor-to-scalar ratio bound} is slightly relaxed, and we find $m \lesssim \num{3.8e-6} \MP$, reflecting that we gained slight flexibility in adapting the inflaton potential at higher scales.
Next, consider how additional backreaction affects these bounds. Backreaction implies that inflation is prolonged without a need for the inflaton to traverse down the potential. This entails that the inflaton amplitude corresponding to $f_{\mathrm{thr}}$ can be lower than in a scenario without backreaction. For example, in a scenario with extreme backreaction, $\Delta N_{\mathrm{BR}} = 20$, the bound on the inflaton mass for $f_{\mathrm{thr}} = \SI{e-6}{\hertz}$ is relaxed; $m \lesssim \num{6e-6} \MP$.
Note, however, that the extended duration of inflation due to backreaction should also not be too large such that it does not affect the physics corresponding to frequencies $f < f_{\mathrm{thr}}$. 
This surely occurs when the term in square brackets in Eq.~\eqref{eq: V threshold} becomes negative.
In this discussion, we only accounted for Eq.~\eqref{eq: tensor-to-scalar ratio bound}, as the non-Gaussianity bound, Eq.~\eqref{eq: non-Gaussianity bound}, only becomes more constraining for $\beta \gtrsim 80$.
Ultimately, we find that the bound on the inflaton mass is not very sensitive to the threshold frequency $f_{\mathrm{thr}}$, thus we opt for the conservative estimate $m \lesssim \num{3.1e-6} \MP$ for $f_{\mathrm{thr}} = f_{_\mathrm{BBN}}$ and $\Delta N_{\mathrm{BR}} = 0$.

We rely on the GEF to solve the dynamics of axion inflation. The details of this method may be found in our previous papers, Refs.~\cite{Gorbar_2021,eckardstein_2023}; for the convenience of the reader, we also summarize the main ingredients of this method in Appendix~\ref{app: GEF}. 
We initialize the GEF assuming all gauge-field correlators are zero and with the inflaton field on the slow-roll attractor, approximately 61 $e$-folds before the expected end of inflation in the absence of backreaction,
$\varphi(0) = 15.55 \MP$ and $\dot{\varphi}(0) = - \sqrt{2 /3} m \MP$.
In order to treat all results on an equal footing, we evolve our system until the end of inflation, which we determine based on $\ddot a < 0$. We perform these computations of the background dynamics using the newly available \emph{Gradient Expansion Formalism Factory (GEFF)}~\cite{vonEckardstein:2025jug}, available at \href{https://github.com/richard-von-eckardstein/GEFF}{https://github.com/richard-von-eckardstein/GEFF}.

\subsection{Gravitational-wave modeling, detection characteristic and constraints}

To compute the SGWB spectrum, we use the evolution of the background fields obtained by the GEF to solve the respective mode equations, Eq.~\eqref{eq: Mode Eq - Pure}.
These mode functions can then be utilized to compute the induced tensor power spectrum, Eq.~\eqref{eq: PT}; for details, see Appendix~\ref{app: PT}.
For all results, we assume that reheating occurs instantaneously. These computations are also performed using the \textsc{GEFF} package~\cite{vonEckardstein:2025jug}.

To estimate the observational prospects for detecting a GW signal with a given GW observatory, we shall employ the signal-to-noise ratio (SNR) defined as~\cite{allen_1996, allen_1997, maggiore_2000}
\begin{equation}
    S/N = \left(n_{\mathrm{det}} t_{\mathrm{obs}} \int_{f_{\mathrm{min}}}^{f_{\mathrm{max}}} \D f \, \left( \frac{\Omega_{\rm signal}(f)}{\Omega_{\rm noise}(f)}\right)^2 \right)^{1/2}\, .
    \label{eq: SNR}
\end{equation}
The integration range $[f_{\mathrm{min}}, f_{\mathrm{max}}]$ corresponds to the frequency band of the detector, $n_{\mathrm{det}} = 1$ or $2$ distinguishes between auto-correlation and cross-correlation searches for an SGWB signal, and $t_{\mathrm{obs}}$ is the observation time.
For experiments that are currently not operational, we set $t_{\mathrm{obs}} = \SI{1e0}{\rm yr}$. The data for the noise spectra, $\Omega_{\rm noise}$, are taken from Refs.~\cite{schmitz_2021}. Specifically, we will consider the detection prospects of ET\footnote{Similar results can be obtained for Cosmic Explorer (CE), whose sensitivity is comparable to ET~\cite{CE_2019}.} with an expected frequency coverage of $\SI{1}{\hertz} -  \SI{10}{\kilo\hertz}$, and the LISA space mission, sensitive to frequencies $\SI{10}{\micro\hertz} -   \SI{1}{\hertz}$~\cite{ET_2010, LISA_2017, LISA_2019}.
We furthermore consider constraints imposed by the non-detection of an SGWB by the LIGO--Virgo detector network after its observing run 3, i.e., HLVO3, which includes observations by the LIGO detectors in Hanford (H) and Livingston (L) as well as observations by the Virgo (V) detector at frequencies $\SI{10}{\hertz} -  \SI{1}{\kilo\hertz}$~\cite{LIGO_2010, LIGO_2014}.

The energy density of GWs contributes to the energy budget of dark radiation, which can be quantified in terms of the effective number of relativistic degrees of freedom in addition to SM photons and neutrinos. We denote the difference between the total effective number of relativistic degrees of freedom, $N_{\mathrm{eff}}$, and the SM prediction~$N_{\mathrm{eff}}^{\mathrm{SM}}$~\cite{drewes_2024} by
\begin{equation}
    \Delta N_{\mathrm{eff}} = N_{\mathrm{eff}} - N_{\mathrm{eff}}^{\mathrm{SM}}  , \qquad N_{\mathrm{eff}}^{\mathrm{SM}} =  3.0440 \pm 0.0002 \, ,
\end{equation}
which allows us to write the upper limit on the allowed amount of dark GW radiation as~\cite{caprini_2018}
\begin{equation}
    \int_{f_{_\mathrm{BBN}}}^{f_{\mathrm{end}}} \frac{\D f}{f} h^2 \Omega_\mathrm{GW}(f) \lesssim \num{5.6e-6} \Delta N_{\mathrm{eff}} \, .
    \label{eq: Neff}
\end{equation}
While PLANCK data alone imposes the limit $\Delta N_{\mathrm{eff}} \lesssim 0.33$~\cite{planck_2020_X}, combined BBN and CMB data suffer from larger uncertainties~\cite{yeh_2021, pisanti_2021}, which weakens this limit. 
Following Ref.~\cite{agazie_nanograv_RPL_2025}, we will therefore work with a conservative limit, $\Delta N_{\mathrm{eff}} \lesssim 0.5$.\footnote{
Ref.~\cite{adshead_2020_A,adshead_2020_B} used $\Delta N_{\mathrm{eff}} \lesssim 0.33$; however, our results are insensitive to this small difference in $\Delta N_{\mathrm{eff}}^{\rm max}$.}


\section{Parameter scan}
\label{sec: Constraints on strong BR}


\subsection{Relevant parameter region}
\label{sec: Parameter space estimates}

To study GW production from axion inflation leading up to the strong-backreaction regime, we first estimate the parameter region in which GW production is not too weak to be feasibly observable and not too strong to obey the bound $\Delta N_{\mathrm{eff}} < 0.5$.

To this end, we employ a similar strategy as in Ref.~\cite{adshead_2020_A}:
Considering that the induced tensor power spectrum in the weak-backreaction regime scales with $H$ and $\xi$ as in Eq.~\eqref{eq: PT analytical}, we can determine how a decrease in the inflaton mass $m \propto H$ needs to be compensated by an increase in the axion--vector coupling $\beta$ to maintain an equal amplitude in the tensor power spectrum. 
Taking into account that $|\xi| = \beta|\p \varphi/\p N|/(2\MP) $, one arrives at%
\footnote{This expression matches with Eq.~(29) in Ref.~\cite{adshead_2020_A} up to a factor of $1/2$.
However, this missing factor is most likely due to a typo in Ref.~\cite{adshead_2020_A}, as it is required to obtain their subsequent equation, Eq.~(30).} 
\begin{equation}
    \log_{10} \left(\frac{m}{\MP} \right) \simeq -\frac{\pi}{2 \log(10)} \left| \frac{1}{\MP} \frac{\p \varphi}{\p N} \right| \beta + C \,,
    \label{eq: scaling estimate}
\end{equation}
with $C$ an arbitrary constant. Interestingly enough, the same scaling relation can be obtained by demanding constant backreaction parameters $\delta_{\mathrm{KG}}, \delta_{\mathrm{F}}$ for decreasing inflaton mass (see Eq.~\eqref{eq: BR parameters}).
During chaotic inflation, we have that $|\p \varphi /\p N| \lesssim 1.4 \MP$. However, in contrast to Ref.~\cite{adshead_2020_A}, wherein the emphasis is on GW production during preheating, we allow for GFIGW production already before the end of inflation.
Therefore, the log-linear slope may be steeper than their estimate, $(\beta_2 - \beta_1) \simeq 1.1 \log_{10} (m_1/m_2)$.
To constrain the constant $C$, we again use Eq.~\eqref{eq: PT analytical}, together with Eq.~\eqref{eq: GW from PT} and Eq.~\eqref{eq: scaling estimate},
\begin{equation}
    \log_{10} (h^2 \Omega_\mathrm{GW}^{\mathrm{min}})^{1/4} + 3.3 \lesssim C \lesssim \log_{10} (h^2 \Omega_\mathrm{GW}^{\mathrm{max}})^{1/4} + 3.3 \, ,
    \label{eq: Constant constraint}
\end{equation}
where $ h^2 \Omega_\mathrm{GW}^{\mathrm{min}}$ and $h^2 \Omega_\mathrm{GW}^{\mathrm{max}}$ are the minimal and maximal amplitudes in between which the SGWB signal is supposed to fall.
To expect any observable signal for LISA or ET, we conservatively estimate $ h^2 \Omega_\mathrm{GW}^{\mathrm{min}} \approx \num{e-15}$, allowing for strong-backreaction effects to enhance the signal after the slow-roll phase.
At the same time, we estimate $h^2 \Omega_\mathrm{GW}^{\mathrm{max}} \approx \num{e-5}$ to ensure $\Delta N_{\mathrm{eff}} < 0.5$. These estimates result in $-0.45 \lesssim C \lesssim 2.05$.

\subsection{Detailed constraints and detectability regions}

\begin{figure}
    \centering
    \includegraphics[width=0.95\textwidth]{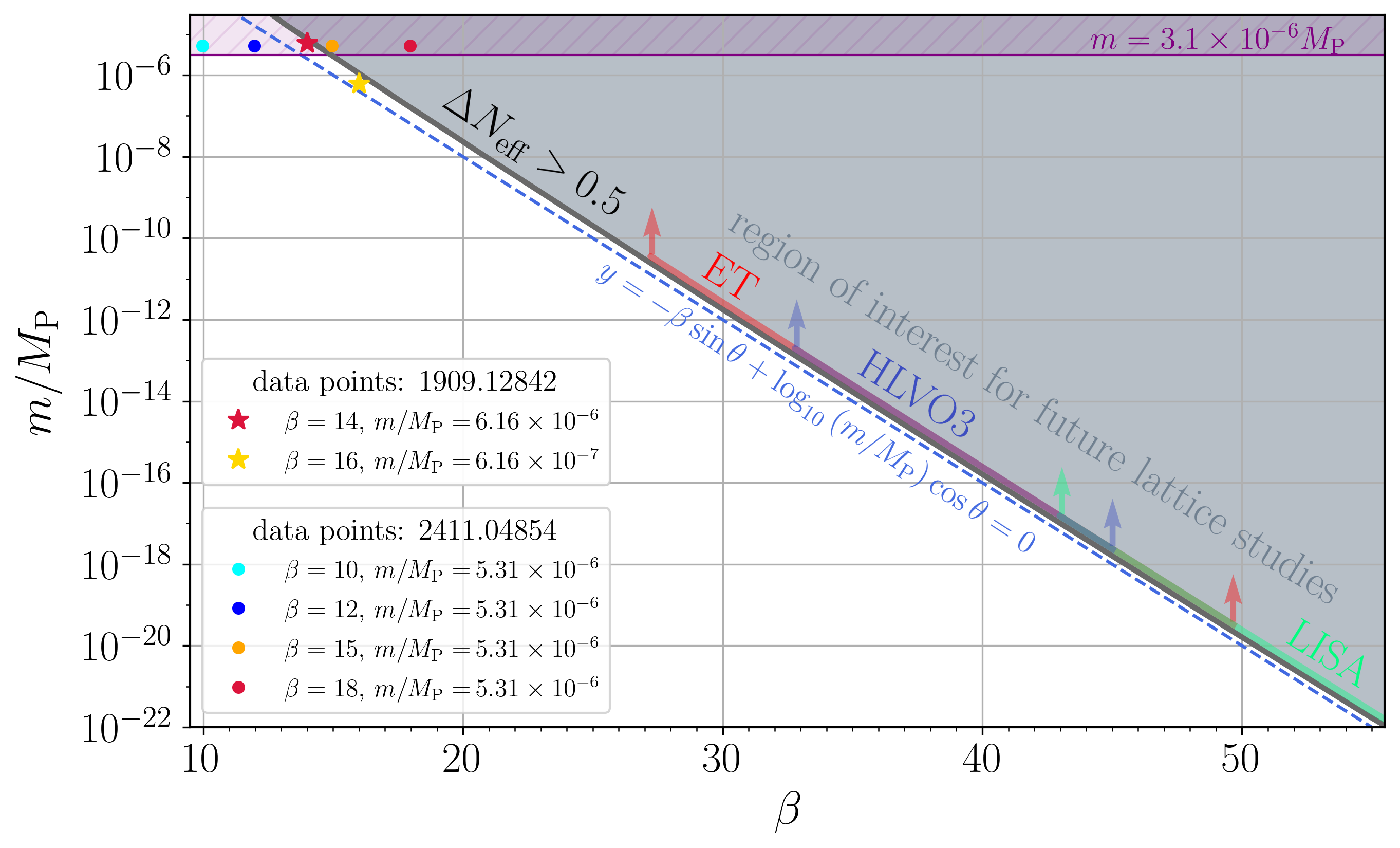}
    \caption
    {
     Results of our final parameter scan shown in terms of $\beta$ and $m/\MP$.
     The gray-shaded area shows the region where $\Delta N_{\mathrm{eff}} > 0.5$, implying an overproduction of GFIGWs in conflict with CMB and BBN constraints.
     The purple-shaded area shows $m > \num{3.1e-6} \MP$, which is excluded due to the bound on the tensor-to-scalar ratio as given by Eq.~\eqref{eq: tensor-to-scalar ratio bound} 
     for a threshold frequency $f_{\mathrm{thr}} = f_{_\mathrm{BBN}}$ and $\Delta N_{\mathrm{BR}} = 0$.
     We also depict regions of $S/N > 1$ for ET in red, LISA in green and HLVO3 in blue.
     The red and yellow stars indicate respectively $\beta = 14,\, m = \num{6.16e-6} \MP$ and $\beta = 16,\, m = \num{6.16e-7} \MP$,
     the approximate parameter points for which Ref.~\cite{adshead_2020_A} found $\Delta N_{\mathrm{eff}} > 0.33$ for gauge-field production during preheating.
     The light-blue, blue, orange, and red points indicate $\beta \simeq 10,\, 12,\, 15$, and $18$ for $m \simeq \num{5.31e-6}$ corresponding to results obtained by Ref.~\cite{sharma_2025}.
     The first two points violate $\Delta N_{\mathrm{eff}}$ bounds only during preheating, the {latter} two avoid $\Delta N_{\mathrm{eff}}$ bounds due to strong backreaction including axion inhomogeneities.
     }
    \label{fig: Parameter Scan beta-m}
\end{figure}

\begin{figure}
    \centering
    \includegraphics[width=0.95\textwidth]{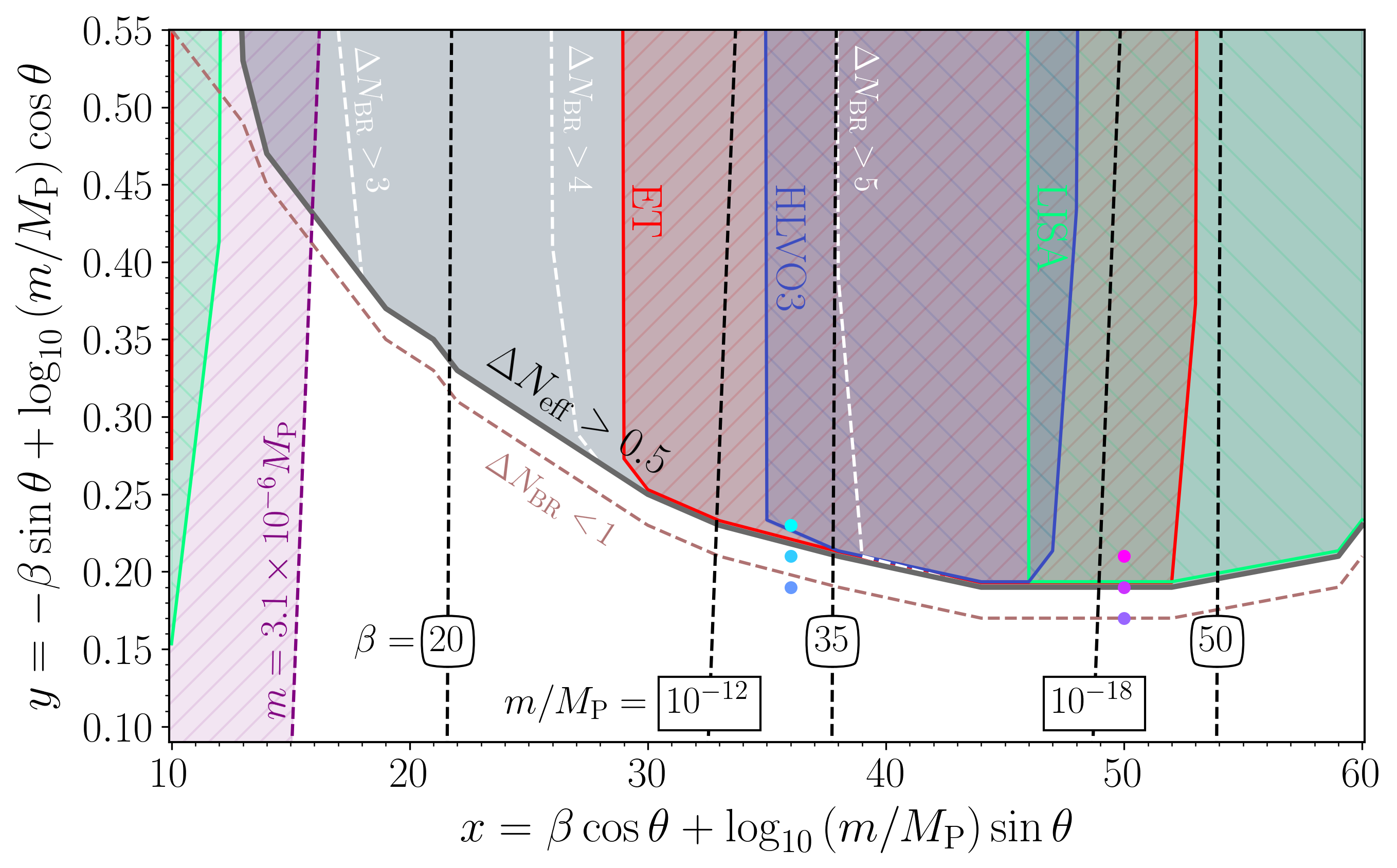}
    \caption
    {
     Same results as in Fig.~\ref{fig: Parameter Scan beta-m} shown in terms of $x$ and $y$. 
     As before, the gray shaded area indicates $\Delta N_{\mathrm{eff}} > 0.5$ and the purple shaded area indicates $m > \num{3.1e-6}$.
     The red, blue, and green shaded areas indicate $S/N > 1$ for ET, HLVO3, and LISA, respectively. Clearly, $S/N > 1$ is only possible for parameter space points in conflict with $\Delta N_{\mathrm{eff}} < 0.5$.
     In white dashed lines, we indicate where the duration of inflation is extended by more than $\Delta N_{\mathrm{BR}} \simeq 3,\, 4,\, 5$ due to strong-backreaction effects.
     In dashed brown, we show where the extended duration of inflation is less than one $e$-fold, $\Delta N_{\mathrm{BR}} < 1$. 
     Evidently, strong backreaction is required for an observable signal, but simultaneously leads to an overproduction of gravitational radiation.
     The points indicated in shades of light-blue are $x=36$ and $y = 0.19,\, 0.21,\, 0.23$. The magenta points are $x=50$ and $y = 0.17,\, 0.19,\, 0.21$.
     SGWB spectra for these six benchmark points are shown in Fig~\ref{fig: BenchmarksPoints}.
     }
    \label{fig: Parameter Scan x-y}
\end{figure}

In a first preliminary scan of the $(\beta, \log_{10} (m/\MP) )$ parameter space, we shall identify the relevant parameter region and power-law dependence expected from the estimate in Eq.~\eqref{eq: scaling estimate}. 
Given the constraints on the inflaton mass, we begin by studying a grid of $21 \times 16$ evenly spaced points for $\beta \in [10, 50]$ and $\log_{10} (m/\MP) \in [-5, -20]$.
From this first analysis, we find $\Delta N_{\mathrm{eff}} > 0.5$ for parameter points following the log-linear relationship $ \log_{10} (m / \MP ) \gtrsim \tan \theta \beta + 0.3$ with $\tan \theta \approx -0.4$,
while finding no data points with $S/N > 1$ for LISA, ET, or HLVO3 that are not simultaneously in conflict with the $\Delta N_{\mathrm{eff}}$ bound.
Motivated by the results of this preliminary analysis, we next perform a follow-up scan in a suitably rotated parameter space given by
\begin{equation}
    \begin{pmatrix}
        x \\
        y
    \end{pmatrix} = 
    \begin{pmatrix}
        \cos \theta & \sin \theta  \\
        -\sin \theta & \cos \theta
    \end{pmatrix}
    \begin{pmatrix}
        \beta \\
        \log_{10} \left( m/\MP \right)
    \end{pmatrix}
\end{equation}
with $\cos \theta \approx 0.93, \sin \theta \approx -0.37$. 
Variations of $y$ thus correspond to varying the constant offset of the line $ \log_{10} (m / \MP) =  \tan \theta \beta + C$, with $y = \cos \theta C \gtrsim -0.44$.
This second parameter scan covers a grid of 31 x 21 points for $x \in [15, 45]$ and $y \in [-1.0, 1.0]$, 
choosing a mildly lower range for $y$ to allow for strong-backreaction effects.
The second scan confirms the previous results, but also indicates a mild drift in the bound away from a purely log--linear relation at higher masses and smaller couplings. 
At this resolution, still no viable region with $S/N > 1$ for LISA, ET, or HLVO3 is found.
Finally, we perform a third scan over $51\times 21$ evenly spaced points over a narrower range in $y$, $ y \in [0.15, 0.55]$ for $x \in [10, 60]$.\footnote{
Technical limitations of our GEF approach prohibit us from scanning down to larger values in $x$ with the same consistent density in parameter points.
This is due to the increasingly prolonged strong-backreaction regime for $\beta \gtrsim 60$.
Therefore, we are not able to cover the entire range of LISA sensitivity down to $m \simeq \num{e-33}\MP$, $\beta \simeq 80$.
PTA sensitivity would require values as low as $m \simeq \num{e-43}\MP$, $\beta \simeq 105$.}
The results of this final, third parameter scan are shown in Figs.~\ref{fig: Parameter Scan beta-m} and \ref{fig: Parameter Scan x-y}.

In Fig.~\ref{fig: Parameter Scan beta-m}, we demonstrate the region excluded by $\Delta N_{\mathrm{eff}} > 0.5$ as a gray shaded area.
We also show the constraints on the inflaton mass which follow from Eqs.~\eqref{eq: tensor-to-scalar ratio bound} and \eqref{eq: non-Gaussianity bound}. Choosing a conservative estimate for this bound, with $f_{\mathrm{thr}} = f_{_\mathrm{BBN}}$ and $\Delta N_{\mathrm{BR}} = 0$, this bound is given by $m \lesssim \num{3.1e-6} \MP$ as indicated by the purple shaded band.
Regions with $S/N > 1$ for ET, LISA and HLVO3 are also depicted in red, green and blue, respectively.
Furthermore, we indicate six additional benchmark points that correspond to the results obtained by two separate lattice studies of PAI, Refs.~\cite{adshead_2020_A, sharma_2025}.
The two points marked with stars are $\beta \simeq 14$, $m \simeq \num{6.16e-6} \MP$ and $\beta \simeq 16$, $m \simeq \num{6.16e-7} \MP$ in red and yellow, respectively.
For these points, Ref.~\cite{adshead_2020_A} found violations of $\Delta N_{\mathrm{eff}} < 0.33$ by accounting for gauge-field production after the end of inflation during preheating.
The $\beta$ values of these points were estimated based on Fig.~1 in Ref.~\cite{adshead_2020_A} and may not be exact.
The four points marked with dots are $\beta \simeq 10,\, 12,\, 15$, and $18$ for $m \simeq \num{5.31e-6} \MP$ marked in light-blue, dark-blue, orange, and red, respectively.
For the first two points, Ref.~\cite{sharma_2025} found $\Delta N_{\mathrm{eff}}$-violation from gauge-field production after inflation.
For the latter two, they instead obtained no $\Delta N_{\mathrm{eff}}$-violation during inflation, which already accounts for strong backreaction including axion inhomogeneities.
These lattice data points clearly underpin that the excluded region, $\Delta N_{\mathrm{eff}} > 0.5$, following from our results is not a hard bound on PAI. 
It merely represents a GEF benchmark which may be used for more intensive follow up lattice studies, which can account for inhomgeneities and preheating effects beyond the validity of homgeneous PAI.
Nevertheless, as we shall discuss in the subsequent paragraphs, our results still allow us to draw important conclusions about the onset of strong backreaction and the detection prospects for GFIGWs from PAI in next generation interferometer experiments.

In Fig.~\ref{fig: Parameter Scan x-y}, we show our results in the rotated $(x,y)$ coordinate plane, which allows us to better resolve several interesting features of our results. 
In dashed black lines, we indicate reference lines of constant $\beta$ and $m /\MP$. 
In this very narrow diagonal slice of the $\beta$--$\log_{10} (m/\MP)$ plane, variations in $y$ along constant $x$ keep $\beta$ roughly constant while only very mildly varying $m /\MP$. 
However, our preliminary scans indicate that such a high degree of precision is necessary to properly resolve the physical effects at play.

As in Fig.~\ref{fig: Parameter Scan beta-m}, we show $\Delta N_{\mathrm{eff}} > 0.5$ in shaded gray and $m > \num{3.1e-6} \MP$ in shaded purple.
We also indicate regions of $S/N > 1$ in shaded red, blue, and green for ET, HLVO3, and LISA, respectively. Strikingly, no regions of $S/N > 1$ may be found outside the region of $\Delta N_{\mathrm{eff}} > 0.5$.
The green and red shaded regions at $x \lesssim 12$ correspond to points for which $\mathcal{P}_T^{\mathrm{vac}}$ would be observable with LISA and ET. These regions are, however, excluded by our CMB bounds.
Furthermore, in dashed white lines, we indicate regions where the duration of inflation is extended by more than $\Delta N_{\mathrm{BR}}$ $e$-folds, a smoking-gun signature of strong backreaction.
Notably, $\Delta N_{\mathrm{BR}} > 3,\, 4,\, 5$ occur when $\Delta N_{\mathrm{eff}} > 0.5$, 
indicating that the extended period of gauge-field production due to strong backreaction leads to an inevitable overproduction of gravitational radiation in conflict with CMB and BBN constraints.
We also indicate the region where $\Delta N_{\mathrm{BR}} < 1$ by a dashed brown contour line. Below this contour, solutions to the EOMs of PAI show no sizable backreaction effects. This region is clearly outside the boundary describing $\Delta N_{\mathrm{eff}} > 0.5$. 
This strengthens the interpretation that the additional gauge-field production in the strong-backreaction regime is responsible for $\Delta N_{\mathrm{eff}}$ violation, 
while also being a prerequisite for a strong enough signal to be observable by current and planned interferometer experiments such as ET, LISA, and HLVO3.
To illustrate this fact, we pick six benchmark points, $x=36,\, y = 0.19,\, 0.21,\, 0.23$ and $x=50,\, y = 0.17,\, 0.19,\, 0.21$, indicated in Fig.~\ref{fig: Parameter Scan x-y} by three light-blue and magenta points, respectively, for which we discuss the dynamical evolution and SGWB spectra in more detail.

\begin{figure}
    \centering
    \includegraphics[width=0.95\textwidth]{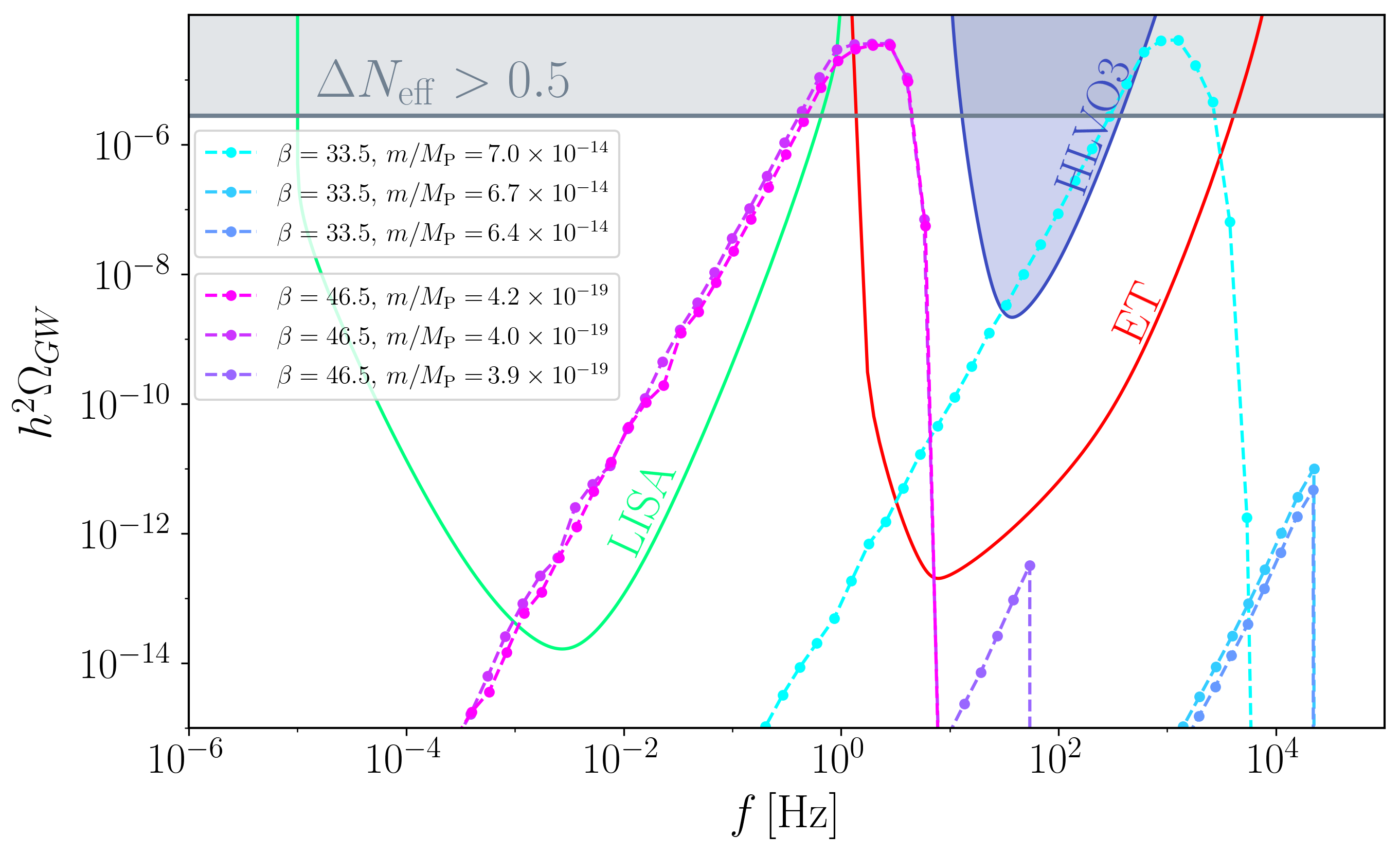}
    \caption
    {
     SGWB spectra for the six benchmark points marked in Fig.~\ref{fig: Parameter Scan x-y}. We explicitly plot our numerical data points and connect them by piecewise linear interpolation.
     Spectra in shades of light-blue correspond to $\beta = 33.5$ and $m = 6.4,\, 6.7,\, \num{7.0e-14} \MP$ or $x=36$ and $y = 0.19,\, 0.21,\, 0.23$, 
     those in shades of magenta correspond to  $\beta = 46.5$ and $m = 3.9,\, 4.0,\, \num{4.2e-19} \MP$ or $x=50$ and $y = 0.17,\, 0.19,\, 0.21$.
     In red, blue, and green, we show power-law integrated sensitivity curves for ET, LISA and HLVO3. In gray, we suggest $\Delta N_{\mathrm{eff}} > 0.5$ (although Eq.~\eqref{eq: Neff} is an integrated bound).
     For each set of three benchmark points, one can notice a qualitative threshold mass, $m \simeq \num{6.7e-14} \MP$ for $\beta = 33.5$ or $m \simeq \num{3.9e-19} \MP$ for $\beta = 46.5$.
     A small increase past this threshold mass triggers strong backreaction, leading to a drastically increased SGWB amplitude, while also redshifting the signal to lower frequencies.
     Without strong backreaction, signals are too weak to be observed, with strong backreaction, the signal conflicts with $\Delta N_{\mathrm{eff}} < 0.5$.
    }
    \label{fig: BenchmarksPoints}
\end{figure}

In Fig.~\ref{fig: BenchmarksPoints}, we show the SGWB spectra for these six benchmark points to illustrate the physical effect behind our results. These are shown with the same color scheme as in Fig.~\ref{fig: Parameter Scan x-y}.
In physical parameters, they correspond to $\beta = 33.5$ and $m = 6.4,\, 6.7,\, \num{7.0e-14} \MP$ in shades of light-blue and $\beta = 46.5$ and $m = 3.9,\, 4.0,\, \num{4.2e-19} \MP$ in shades of magenta.
For both fixed values of $\beta$, no strong backreaction occurs below a certain threshold mass, $m \simeq \num{6.7e-14} \MP$ for $\beta = 33.5$ or $m \simeq \num{3.9e-19} \MP$ for $\beta = 46.5$.
The gauge-field-induced friction on the inflaton is just small enough such that the inflaton velocity becomes comparable to the inflaton potential energy, $\dot{\varphi}^2 \sim V(\varphi)$, thus ending inflation on the usual slow-roll trajectory.
However, only a very mild increase past this threshold mass allows a marginally larger amount of gauge-field production, slightly prolonging inflation, such that the system has just enough time to enter the strong-backreaction regime.
Once this process is triggered, inflation is delayed by multiple $e$-folds as the inflaton kinetic energy oscillates while the energy density in the electromagnetic field further increases sourcing additional GWs. 
In this case, inflation only ends once $\rho_{\mathrm{EM}} \simeq V(\varphi)$.
Ultimately, we find that the signal in the weak-backreaction regime is too low to reach the sensitivity of current or planned interferometers, while too much gravitational radiation from late time strong backreaction leads to a conflict with CMB and BBN bounds on $\Delta N_{\mathrm{eff}}$.

\begin{figure}
    \centering
    \includegraphics[width=0.95\textwidth]{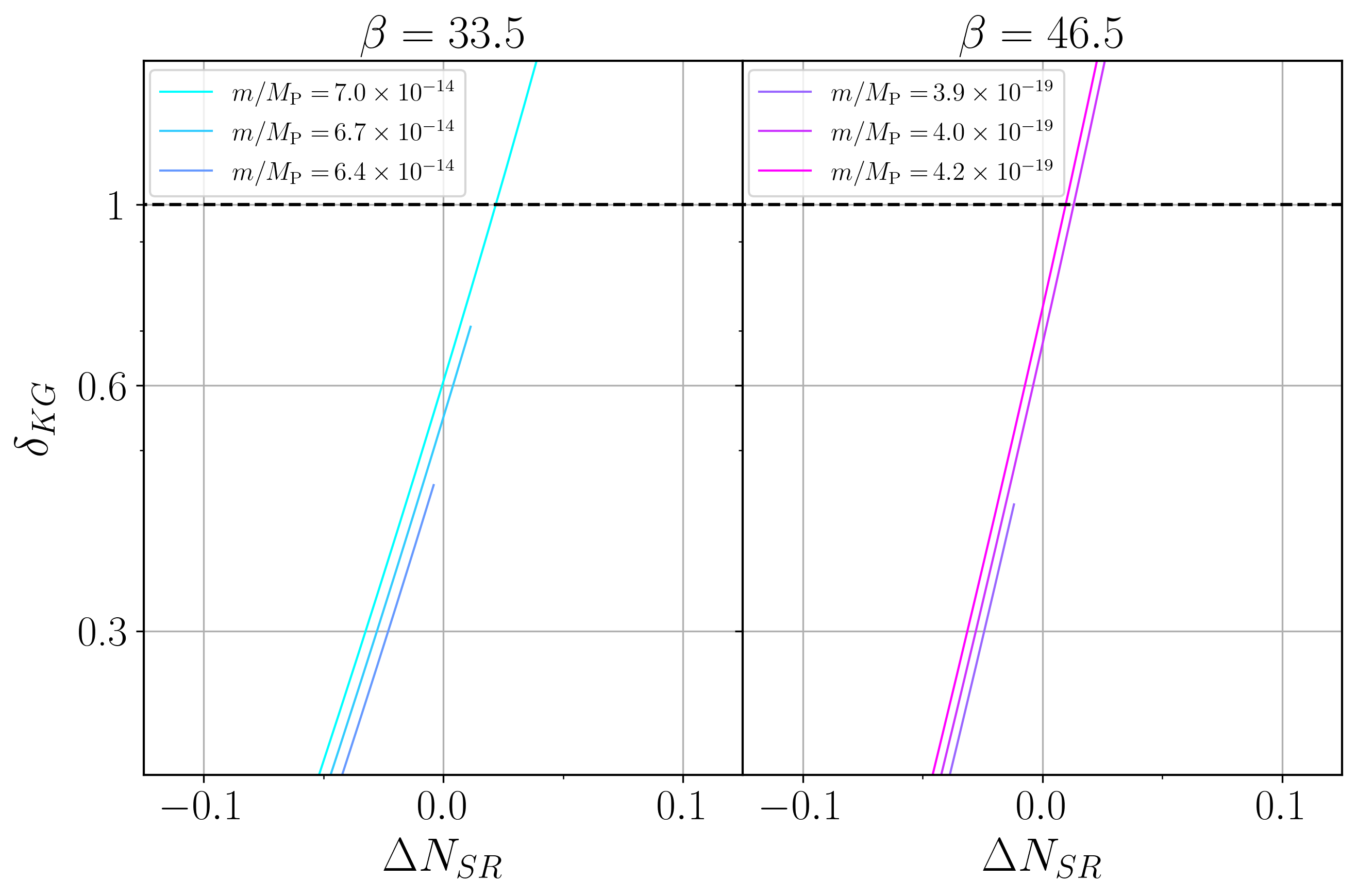}
    \caption
    {
     Evolution of the backreaction parameter $\delta_{\mathrm{KG}}$ for the six benchmark points shown in Fig.~\ref{fig: BenchmarksPoints} with the corresponding color scheme.
     On the horizontal axis, we show the number of $e$-folds $\Delta N_{\mathrm{SR}}$ after the naive end of slow-roll inflation, $N \simeq 61$ $e$-folds after initializing our system.
     Notably, for the benchmark points with $\Delta N_{\mathrm{eff}} < 0.5$, inflation ends just before $\delta_{\mathrm{KG}} = 1$.
    }
    \label{fig: BR Parameter}
\end{figure}

To further illustrate this point, we show the evolution of the backreaction parameter $\delta_{\mathrm{KG}}$ just before the end of slow-roll inflation in Fig.~\ref{fig: BR Parameter} for the same parameter space points as in Fig.~\ref{fig: BenchmarksPoints} and using the same color scheme as before.
For the three parameter points that manage to avoid $\Delta N_{\mathrm{eff}} > 0.5$, inflation ends on the slow-roll attractor, $\Delta N_{\mathrm{SR}} \simeq 0$, just before  $\delta_{\mathrm{KG}} = 1$, thus never triggering strong backreaction. 
This is due to two effects: since $\delta_{\mathrm{KG}} \propto H^2 \propto m^2$, it is marginally larger for larger masses. 
Simultaneously, inflation is marginally prolonged for larger masses, possibly due to small corrections in the Friedmann equation coming from $\delta_{\mathrm{F}} \propto H^2 \propto m^2$.
In fact, the second backreaction parameter, $\delta_{\mathrm{F}}$, is smaller than $\delta_{\mathrm{KG}}$ by about one order of magnitude at $\Delta N_{\mathrm{SR}} \simeq 0$.
This implies that $\rho_{\mathrm{EM}}$ already takes a subdominant, yet non-negligible, fraction of the total energy density, indeed leading to small extensions of the duration of inflation.
Combined, these two effects explain how only a small increase in the inflaton mass allows the system to reach $\delta_{\mathrm{KG}} = 1$. thereby triggering strong backreaction.

Returning to the $\Delta N_{\mathrm{eff}} = 0.5$ contour line, we perform a phenomenological fit to the boundary of the region shown in Fig.~\ref{fig: Parameter Scan x-y}. Motivated by the approximate scaling relation between $\beta$ and $\log_{10} (m/\MP)$ in Eq~\eqref{eq: scaling estimate}, we describe the viable parameter region by
\begin{equation}
    \log_{10} \left( \frac{m}{\MP} \right) \lesssim t_1 \beta +  t_2 \log_{10} \left(\frac{\beta}{10} \right) + C,
    \label{eq: Exclusion}
\end{equation}
allowing for a mild deviation from a linear relation to capture logarithmic terms neglected in the derivation of Eq.~\eqref{eq: scaling estimate}.
Our fit yields the coefficients $t_1 = -0.387$, $t_2 = -1.34$, and $C = 0.51$. Note that $t_1$ is consistent with the initial parameter scan, $\tan \theta \simeq -0.4 \simeq t_1$.
From Eq.~\eqref{eq: scaling estimate}, we can infer that this slope roughly corresponds to $|\p \varphi/ \p N| = 0.6 \MP$, which occurs at $\Delta N_{\mathrm{SR}} \simeq 3 - 4$ on the slow-roll attractor of chaotic inflation.
Notably, from the analytical estimate in Eq.~\eqref{eq: BR parameters}, this implies $\delta_{\mathrm{KG}} \sim \num{e-3}$, thus backreaction is still negligible at this moment in time.
A possible explanation for the significance of this time scale is the retarded response in the gauge-field production following a change in the inflaton velocity.
If inflation lasts slightly longer, the retarded production of gauge fields sets in just in time to trigger strong backreaction, preventing the immediate end of inflation.
That this effect is correlated with the production of GWs is deduced from the fact that both backreaction and tensor perturbations scale according to Eq.~\eqref{eq: scaling estimate} (cf.\ Eqs.~\eqref{eq: PT analytical} and \eqref{eq: BR parameters}).

From the results obtained by Refs.~\cite{adshead_2020_A,adshead_2020_B,sharma_2025}, which we indicate in Fig.~\ref{fig: Parameter Scan beta-m},
it should be clear that the contour in Eq.~\eqref{eq: Exclusion} is not a hard exclusion limit.
Our results are limited by our computational method, as we are not able to include axion inhomogeneities nor can we consistently follow gauge-field production into the post-inflationary epoch to study the effects of preheating.
References~\cite{adshead_2020_A,adshead_2020_B} have shown that, for $\beta = 14,\, m=\num{6.16e-6} \MP$ and $\beta = 16,\, m=\num{6.16e-7} \MP$, gauge-field production leads to a post-inflationary violation of $\Delta N_{\mathrm{eff}}$. 
The same was found by Ref.~\cite{sharma_2025} for $\beta = 10,\, 12$ and $m=\num{5.31e-6} \MP$.
In all these four cases, their parameter points lie below our excluded region.
From this observation, we may deduce that an absence of gauge-field backreaction during inflation does not necessarily imply that gauge-field dynamics cannot be enhanced by such effects after inflation has ended. 
Our results, however, appear consistent within the scope of our method: 
For $\beta = 10,\, 12$ and $m=\num{5.31e-6} \MP$, the benchmark points studied in Ref.~\cite{sharma_2025}, the authors find that strong backreaction is not reached during inflation, which is in agreement with our results.
This accordance is consistent with previous observations on the lattice~\cite{Figueroa_2023, Figueroa_2024} demonstrating that GEF results accurately capture the weak-backreaction regime as well as the onset of strong backreaction.
In summary, the first major result of our analysis is consistent in the lead up to the strong-backreaction regime: observable GW signals from Abelian PAI within the reach of current or near-future interferometers can only be attained for parameter values that also result in strong backreaction.

Meanwhile, the onset of strong backreaction may not necessarily imply an overproduction of gauge-field-sourced gravitational radiation. 
Comparing again to the lattice simulation performed in Ref.~\cite{sharma_2025}, we find that their data points $\beta = 15,\, 18$ and $m=\num{5.31e-6} \MP$ are also consistent with an extended period of inflation due to backreaction effect, while avoiding $\Delta N_{\mathrm{eff}} > 0.5$, as the coupling of gauge fields to axion gradients dampens the GW production. In this way, their results manage to penetrate our excluded region indicated by Eq.~\eqref{eq: Exclusion}.
However, given the dynamical range covered in their simulations, it is not evident if accounting for the dynamics after inflation may lead to an overproduction of GFIGWs, after all. Indeed, the authors of Ref.~\cite{sharma_2025} comment on the model dependence of their results: If the gauge field should quickly decay after inflation because of its interaction with charge carriers (which are not part of the particle spectrum of PAI by definition), reheating may be more or less instantaneous and the bound on $\Delta N_{\rm eff}$ may be avoided. An extended stage of preheating and reheating, on the other hand, may entail the build-up of a large GW energy density that ultimately does violate the upper bound on  $\Delta N_{\rm eff}$.

These caveats notwithstanding, the relevance of the contour line in Eq.~\eqref{eq: Exclusion} should not be understated.
Unlike the lattice results discussed above, the GEF allows us to perform an extensive parameter scan of the PAI parameter space, which allows us to draw conclusions beyond the study of specific benchmark points.
The exclusion region in Eq.~\eqref{eq: Exclusion} represents a GEF benchmark that can be used to guide follow-up lattice studies, indicating regions of interest wherein potentially observable signals for current or future GW observatories may lie.
We summarize these regions in Table~\ref{tab: ROIs}. The ranges indicated in this table should be understood as lying above the contour given by Eq.~\eqref{eq: Exclusion} (see also Fig.~\ref{fig: Parameter Scan beta-m}).
For homogeneous backreaction, we find that all regions of interest within the range of our parameter scan are in conflict with $\Delta N_{\mathrm{eff}} > 0.5$. 
Solving the full dynamics of inhomogeneous backreaction, parts of these regions may yield observable (excludable in the case of HLVO3) signals.
For LISA and NANOGrav, we extrapolate the regions of interest based on Eq.~\eqref{eq: Exclusion} and the frequency corresponding to the end of inflation estimated in Eq.~\eqref{eq: f_end}.

\begin{table}
 \caption{
    Regions of interest for follow-up lattice studies including inhomogeneous backreaction. 
    }
    \begin{center}
    \begin{tabular}{|l|cc|cc|}
        \hline
        \multirow{2}*{\bf Detector} & \multicolumn{2}{c|}{\bf Parameter scan} & \multicolumn{2}{c|}{\bf Extrapolation } \\
        \cline{2-5}
            & $\beta$ & $m/\MP$ & $\beta$ & $m/\MP$ \\
        \hline
        HLVO3 & $ 32.5 - 45.0 $ & $ \num{3e-18} - \num{2e-13}$ & \multicolumn{2}{c|}{---} \\
        ET & $ 27.0 - 49.5 $ & $\num{5e-20} - \num{5e-11}$ &  \multicolumn{2}{c|}{---} \\
        LISA & $42.5 - 56.0$ & $\num{8e-23} - \num{2e-17}$ & $\lesssim 82.0$ & $ \gtrsim \num{5e-33}$ \\
        NANOGrav & \multicolumn{2}{c|}{---} & $\lesssim 107.0$ & $ \gtrsim \num{5e-43}$ \\
        \hline
    \end{tabular}
    \label{tab: ROIs}
    \end{center}
\end{table}
%


\section{Conclusions}
\label{sec: conclusions}


Axion inflation is a fascinating extension of the standard slow-roll inflationary model allowing for a shift-symmetric interaction with gauge fields.
This gauge-field--inflaton coupling leads to the ample production of maximally helical gauge fields due to the spontaneous parity violation associated with the non-zero inflaton velocity during inflation.
At the same time, the gauge fields can backreact onto the inflaton dynamics triggering a regime of non-linear dynamics called the strong-backreaction regime.
Phenomenological consequences of the gauge-field production are numerous, including implications for magnetogenesis, baryogenesis, primordial black hole formation, and GW production.

In this work, we considered the production of GWs sourced by Abelian gauge fields during \textit{pure axion inflation} (PAI), i.e., we studied the production of gauge-field-induced GWs (GFIGWs) in scenarios of axion inflation coupled to a pure Abelian gauge sector. In doing so, we focused on the parametric regime close to the onset of strong backreaction. We investigated in particular the observational prospects for next-generation GW interferometers, i.e., Einstein Telescope (ET) and the Laser Interferometer Space Antenna (LISA), which led us to the conclusion that an observable GW signal from Abelian PAI can only be realized in regions of parameter space that also lead to strong backreaction. 

At the same time, the non-linear dynamics in the strong-backreaction regime result in an extension of inflation past the expected slow-roll stage, allowing for such abundant gauge-field generation that the resulting induced GWs end up being in conflict with BBN and CMB bounds on the number of additional relativistic species in the early Universe, $\Delta N_{\mathrm{eff}}$. We studied this effect in a quadratic inflation potential, arguing that our results are in fact applicable to any inflation model in which the final stage of slow-roll inflation occurs in an approximately quadratic potential.%
\footnote{{However, our conclusions may not hold if the inflaton potential is not approximately parabolic near its minimum or if its shape significantly deviates from a quadratic form at times when the tensor modes relevant for GW observations by LISA and/or ET cross the Hubble horizon; see, e.g., Refs.~\cite{domcke_2016,garcia-bellido_2016,garcia-bellido_2024}. Moreover, the dynamics could differ entirely in multi-field realizations of axion inflation, e.g., in the axion spectator model~\cite{garcia-bellido_2016,ozsoy_2020,unal_2023}.}}
For these models, we find a sharp transition between the regimes of weak and strong backreaction, with slight relative variations in the inflaton mass, $\delta m/m \simeq 0.02$, deciding the fate of the system's evolution:
the end of inflation on the slow-roll trajectory without a detectable signal in ET or LISA or an extended period of strong backreaction leading to a violation of $\Delta N_{\mathrm{eff}} \lesssim 0.5$.

We also worked out a phenomenological parametrization of the phase boundary in parameter space beyond which we observe the violation of the upper bound on $\Delta N_{\mathrm{eff}}$,
\begin{equation}
    \log_{10} \left( \frac{m}{\MP} \right) \simeq t_1 \beta +  t_2 \log_{10} \left( \frac{\beta}{10} \right) + C\,,
\end{equation}
%
with the best-fit values of the coefficients being $t_1 = -0.387$, $t_2 = -1.34$, and $C = 0.51$. The fact that the threshold of strong backreaction coincides with a threshold on the GW amplitude may be explained by considering the similar scaling of the induced power spectrum $\mathcal{P}_T^{\mathrm{ind}}$ and the backreaction parameters $\delta_{\mathrm{KG}}, \delta_{\mathrm{F}}$ (\textit{c.f.} Eqs.~\eqref{eq: PT analytical} and \eqref{eq: BR parameters}).

For the analysis in this paper, we relied on the simplifying assumption of a homogeneous axion field even into the strong-backreaction regime, which allowed us to solve the non-linear dynamics of axion inflation using the gradient expansion formalism (GEF).
Recently, lattice simulations have demonstrated that axion gradients can be important when modeling the gauge-field--inflaton dynamics during strong backreaction~\cite{Figueroa_2023,Figueroa_2024, sharma_2025}.
However, qualitatively, results from the lattice have exhibited the same behavior during strong backreaction: an extended period of inflation during which the gauge-field energy density dominates over the axion kinetic energy.
Furthermore, it is known that the GEF and lattice simulations agree well in the weak-backreaction regime.
Consequently, we expect that the contour line indicated above yields a reasonable estimate for the true results that may be obtained using lattice simulations.
Based on the results by Ref.~\cite{sharma_2025}, it seems likely that the additional impact of axion gradients allows for a smoother transition between the weak and strong-backreaction regime.
This smoother transition may open up the possibility for a detectable GW signal in ET, LISA or even NANOGrav.
In this case, our results serve as a guide for identifying the relevant regions in parameter space for further explorations using lattice techniques, as summarized in Table~\ref{tab: ROIs}.
This also highlights the importance of homogeneous backreaction techniques such as the GEF:
due to their relatively cheap numerical cost and larger dynamical range, they allow for the exploration of large regions of parameter space of axion inflation to guide follow-up lattice simulations.

Finally, we want to comment on the implications of our results for the model of \textit{fermionic axion inflation} (FAI), i.e., a model where the gauge field during axion inflation couples to fermions, resulting in their generation via the Schwinger mechanism. We find that the potential for detecting GWs in the model of PAI, considered in this work, are limited due to the violent response of the inflaton field to the abundant gauge-field production.
Meanwhile, the generation of charge carriers in FAI dampens gauge-field production, which in turn leads to a dampened production of GFIGWs, with important implications for the prospects of a detectable signal in ET, LISA or even PTAs. In the companion paper~\cite{eckardstein_2025a}, we study this intriguing scenario in detail. 


\vskip.25cm
\section*{Acknowledgements}

K.\,S.\ is an affiliate member of the Kavli Institute for the Physics and Mathematics of the Universe (Kavli IPMU) at the University of Tokyo and as such supported by the World Premier International Research Center Initiative (WPI), MEXT, Japan (Kavli IPMU). O.\,S.\ was supported by a Philipp Schwartz fellowship of the University of M\"{u}nster. {At the final stage of this project, the work of O.\,S.\ has received funding through the SAFE\,---\,Supporting At-Risk Researchers with Fellowships in Europe project, which is funded by the European Union. Views and opinions expressed are, however, those of the authors only and do not necessarily reflect those of the European Union, the European Research Executive Agency (REA). Neither the European Union nor the granting authority can be held responsible for them.} Part of this work was conducted using the High Performance Computing Cluster PALMA II at the University of M\"unster~(\url{https://www.uni-muenster.de/IT/HPC})


\appendix
\section{Numerical methods}
\label{app: numerics}

\subsection{Gradient expansion formalism}
\label{app: GEF}
We solve the background dynamics of homogeneous axion inflation via the method known as gradient expansion formalism (GEF). 
This method consists in reformulating the dynamical evolution of our system, Eqs.~\eqref{eq: EoMs}, in terms of bilinear expectation values of electric and magnetic field operators
\begin{subequations}
 \begin{align}
    \GEF{E}{n}&\equiv \frac{a^4}{k_{\mathrm{h}}^{n+4}}\langle \bm{E} \cdot \rot^n \bm{E}\rangle = \int\limits_{0}^{k_{\mathrm{h}}(t)}\frac{\D k}{k} \frac{a^2 k^{n+3}}{2 \pi^2 k_{\mathrm{h}}^{n+4}}  \sum_{\lambda}\lambda^n |\dot{A}_\lambda(t,k)|^2\, ,\\
    \GEF{G}{n}&\equiv -\frac{a^4}{2 k_{\mathrm{h}}^{n+4}}\langle \bm{E} \cdot \rot^n \bm{B} + \bm{B} \cdot \rot^n \bm{E}\rangle \\ 
                    &= \int\limits_0^{k_{\mathrm{h}}(t)} \frac{\D k}{k} \frac{a k^{n+4}}{2 \pi^2 k_{\mathrm{h}}^{n+4}}\sum_{\lambda}\lambda^{n+1} \operatorname{Re}[\dot{A}_\lambda(t,k)A_\lambda^*(t,k)]\, , \nonumber \\
    \GEF{B}{n}&\equiv \frac{a^4}{k_{\mathrm{h}}^{n+4}}\langle \bm{B} \cdot \rot^n \bm{B}\rangle = \int\limits_{0}^{k_{\mathrm{h}}(t)} \frac{\D  k}{k}  \frac{k^{n+5}}{2 \pi^{2}k_{\mathrm{h}}^{n+4}} \sum_{\lambda}\lambda^n |A_\lambda(t,k)|^2 \, ,
 \end{align}
\label{eq: GEF Bilinears}%
\end{subequations}
with $k_{\mathrm{h}}(t) = \underset{s \leq t}{\max} \left\{2|\xi(s)|a(s)H(s)\right\}$, and the solutions to the mode equation~\eqref{eq: Mode Eq - Pure}, $A_\lambda(t,k)$.
The evolution of these quantities is given by an infinite tower of differential equations,
\begin{subequations}
    \begin{equation}
        \dGEF{E}{n} + (4+n)\frac{\D \ln k_{\mathrm{h}}}{\D t} \GEF{E}{n}  + 2\frac{k_{\mathrm{h}}}{a}\GEF{G}{n+1} - 2 I_{,\phi}\dot{\varphi} \GEF{G}{n}  =  S_{\mathcal{E}}^{(n)}\, ,\label{eq: GEF - En}
    \end{equation}
    \begin{equation}
        \dGEF{G}{n} + (4+n)\frac{\D \ln k_{\mathrm{h}}}{\D t} \GEF{G}{n} - \frac{k_{\mathrm{h}}}{a}\left(\GEF{E}{n+1} - \GEF{B}{n+1}\right) - I_{,\phi}\dot{\varphi} \GEF{B}{n} = S_{\mathcal{G}}^{(n)}\, , \label{eq: GEF - Gn}
    \end{equation}
    \begin{equation}
        \dGEF{B}{n} + (4+n)\frac{\D \ln k_{\mathrm{h}}}{\D t} \GEF{B}{n} - 2\frac{k_{\mathrm{h}}}{a}\GEF{G}{n+1}  =  S_{\mathcal{B}}^{(n)}\, . \label{eq: GEF - Bn}
    \end{equation}
    \label{eq: GEF}%
\end{subequations}
The source terms on the right-hand side arise due to the time-varying ultraviolet cut-off, $k_{\mathrm{h}}(t)$, in Eq.~\eqref{eq: GEF Bilinears}. 
Using the approximate analytical solution of Eq.~\eqref{eq: Mode Eq - Pure} in terms of Whittaker functions, $W_{\kappa,\mu}(x)$, these are given by
\begin{subequations}
\begin{align}
    \label{eq: boundary-En}
    S_{\mathcal{E}}^{(n)}&=\frac{d \ln k_{\mathrm{h}}}{d t}\frac{1}{16\pi^{2} |\xi|^2} \sum_{\lambda=\pm 1}\lambda^{n} e^{\lambda \pi \xi} \left| (2i |\xi| - i \lambda \xi ) W_{-i \lambda \xi, \frac{1}{2}}(-4 i |\xi|) + W_{1-i \lambda \xi, \frac{1}{2}}(-4 i |\xi|) \right|^2 \, ,\\
    \label{eq: boundary-Gn}
    S_{\mathcal{G}}^{(n)}&=\frac{d \ln k_{\mathrm{h}}}{d t}\frac{1}{8\pi^{2} |\xi|}\sum_{\lambda=\pm 1}\lambda^{n+1}  e^{\lambda \pi \xi} \operatorname{Re}\left\{W_{1-i \lambda \xi, \frac{1}{2}}(-4 i |\xi|) W_{i \lambda \xi, \frac{1}{2}}(4 i |\xi|)\right\} \, ,\\
    \label{eq: boundary-Bn}
    S_{\mathcal{B}}^{(n)}&=\frac{d \ln k_{\mathrm{h}}}{d t}\frac{1}{4\pi^{2}}\sum_{\lambda=\pm 1}\lambda^{n} e^{\lambda \pi \xi} \left| W_{-i \lambda \xi, \frac{1}{2}}(-4 i |\xi|) \right|^2\, .
\end{align}
\label{eq: Source Terms}%
\end{subequations}
The infinite tower of coupled differential equations~\eqref{eq: GEF} is approximately closed at an order $n_{\mathrm{tr}} \sim \mathcal{O}(100)$ by~\cite{domcke_2024}
\begin{equation}
    \GEF{X}{n_{\mathrm{tr}}+1} \simeq \sum_{l=1}^{L} (-1)^{(l-1)} \left(
    \begin{array}{c}
      L \\
      l
    \end{array}
  \right) 
  \GEF{X}{n_{\mathrm{tr}} - 2l + 1} \,, \quad \mathcal{X} = \mathcal{E},\, \mathcal{G},\, \mathcal{B} \,.
\end{equation}

We solve these equations alongside the Klein--Gordon equation, Eq.~\eqref{eq: phiEoM}, and the first Friedmann equation, Eq.~\eqref{eq: Friedmann}, to obtain the evolution of all background quantities.
To ensure that the end of inflation is reached, our solver checks for sign changes in $\ddot a$. Simultaneously, we ensure that $\rho_{\mathrm{EM}} \geq 0$, the violation of which indicates a clear breakdown of our GEF approximation. By interpolating the time evolution of $\xi(t),\, H(t)$ and $a(t)$ obtained from the GEF, we then solve the mode equation, Eq.~\eqref{eq: Mode Eq - Pure}, for $n_{\mathrm{modes}} = 500$ $k$-values between $k_{\mathrm{IR}} = 10^4 k_{\mathrm{h}}(0)$ and $k_{\mathrm{max}} = 10 k_{\mathrm{h}}(t_{\mathrm{max}})$ by initializing each mode function in the Bunch--Davies vacuum at $k = 10^{5/2} k_{\mathrm{h}}(t_k) \gg k_{\mathrm{h}}(t_k)$ for a given $k$. To assess the GEF solution, we use this mode-by-mode solution (MbM) to compute the relative deviation,
\begin{equation}
    \epsilon_\mathcal{X} = \left|1 - \frac{(\GEF{X}{0})_{\mathrm{GEF}}}{(\GEF{X}{0})_{\mathrm{MbM}}}\right|\, ,
\end{equation}
for $\mathcal{X} = \mathcal{E},\, \mathcal{G},\, \mathcal{B}$, where $(\GEF{X}{0})_{\mathrm{GEF}}$ and $(\GEF{X}{0})_{\mathrm{MbM}}$ are respectively computed from the GEF or the MbM solution.
We demand that the final error, $\epsilon_\mathcal{X}(t_{\mathrm{end}})$, and the RMS error over all $N$ data points, $\epsilon_\mathcal{X}^{\mathrm{RMS}} = \sqrt{\sum_i (\epsilon_\mathcal{X}(t_i))^2/N}$, remain below \SI{10}{\percent} to claim a converged GEF solution.\footnote{
We use the RMS error to mitigate the impact of sharp local minima and maxima in $(\GEF{X}{0})_{\mathrm{GEF}}$ during the fast oscillations associated with the strong-backreaction regime. These peaks can not be accurately resolved by the MbM solution due to the finite resolution in $k$. To avoid discarding well-converged GEF solutions on the basis of these numerical uncertainties, we set the error threshold to \SI{10}{\percent}. However in practice, the final error $\epsilon_\mathcal{X}(t_{\mathrm{end}})$ stayed below \SI{2}{\percent} for nearly all GEF solutions.
}
In case of a larger deviation between the GEF and the MbM, we revert the GEF solution to a state where $\epsilon_\mathcal{X}(t_i) < 0.025$. We then re-initialize the GEF at $t = t_i$ by computing $\GEF{X}{n>0}$ from the MbM solution and repeating the above error estimation until the GEF run has converged. This process typically converges after less than four iterations.

\subsection{Computation of the induced tensor power spectrum}
\label{app: PT}
We compute the vacuum and induced tensor power spectrum in Eq.~\eqref{eq: PT} from the GEF and MbM solution as follows. To solve the dynamics of $u_\lambda^0(\eta, k)$ given by Eq.~\eqref{eq: GW vacuum mode EoM}, we introduce rescaled dimensionless quantities
\begin{equation}
    f_k(\eta) \equiv \sqrt{2 k} a u^0_\lambda(\eta, k),  \qquad g_k(\eta) \equiv \sqrt{\frac{2}{k}} a {u_\lambda^0}'(\eta, k)\, ,
\end{equation}
where we drop the dependence on $\lambda$ since the vacuum modes are unpolarized. These rescaled variables obey the differential equations
\begin{equation}
    \dot{f}_k = \frac{k}{a} g_k + H f_k, \qquad \dot{g}_k = -\frac{k}{a} f_k - H g_k \, ,
\end{equation}
as can be derived from Eq.~\eqref{eq: GW vacuum mode EoM}. 
We initialize these functions deep inside the horizon, $k \gg aH$ as $f_k \to \exp(-i\eta k), \, g_k \to -i\exp(-i\eta k)$, which is equivalent to initializing the GW modes in the Bunch--Davies vacuum.
Explicitly, we impose these initial conditions at $t_k$ defined through $k = 10^{5/2}a(t_k)H(t_k)$ for a given mode $k$.
Expressed in terms of these solutions, the vacuum tensor power spectrum is simply given by
\begin{equation}
    \mathcal{P}_{T,\lambda}^{\mathrm{vac}} (k, \eta) = \frac{2 k^2}{\pi^2 \MP^2} |f_k|^2 \, .
\end{equation}

To compute the induced tensor power spectrum, we need to evaluate the Green function for the differential operator $\mathcal{D}_k$ implicitly defined in Eq.~\eqref{eq: GW vacuum mode EoM}. 
This Green function is given in Eq.~\eqref{eq: Green function}, and can in principle be computed in terms of $f_k$ and $g_k$. 
In practice, however, the term in the denominator in Eq.~\eqref{eq: Green function} suffers from numerical cancellations as $\eta \to \tau$. To mitigate this effect, we instead evolve the Green function itself in time. We define
\begin{equation}
    B_\tau^k(\eta) \equiv 2 k a(\eta)^2 \operatorname{Im} \left[ {u_\lambda^0}^*(\tau, k) \, u_\lambda^0(\eta,k) \right], \quad C_\tau^k(\eta) \equiv 2 a(\eta)^2 \operatorname{Im} \left[ {u_\lambda^0}^*(\tau, k) \, {u_\lambda^0}'(\eta,k) \right] \, ,
\end{equation}
which obey
\begin{equation}
    \dot{B}_\tau^k = \frac{k}{a} C_\tau^k + 2H B_\tau^k \, , \qquad \dot{C}_\tau^k = -\frac{k}{a} B_\tau^k \, ,
    \label{eq: Green function evolution}
\end{equation}
where the time derivative should be understood as acting on the argument $\eta(t)$ and not the subscript $\tau$, which is fixed.
These equations can be solved backwards in time by noting that $B_\tau^k(\tau) = 0$. To derive the second initial condition, $C_\tau^k(\tau)$, note that
\begin{equation}
    W_k(\eta) \equiv \operatorname{Im} \left[ {u_\lambda^0}(\eta, k) \, {{u_\lambda^0}^*}'(\eta,k) \right] = \frac{W_k(\eta_0) a^2(\eta_0)}{a^2(\eta)} = \frac{1}{2 a^2(\eta)} \,,
    \label{eq: Wronskian}
\end{equation}
where $W_k(\eta)$ is, up to a factor of $2i$, the Wronskian of the set of linearly independent solutions to the differential operator $\mathcal{D}_k$.
The last equality in Eq.~\eqref{eq: Wronskian} is derived from the Bunch--Davies vacuum, $a(\eta_0)u_\lambda^0(k,\eta_0) \simeq \exp(-i\eta_0 k)$ for $k \gg \mathcal{H}(\eta_0)$.
Therefore, $C_\tau^k(\tau) = -2 a(\tau)^2W_k(\tau) = -1\, $.
We solve Eq.~\eqref{eq: Green function evolution} backwards in time for a fixed $\eta$ until $t_k$, from whence on we use that
\begin{equation}
    B_\eta^k(\tau) = \frac{ a(\tau)}{a(\eta)} \operatorname{Im} \left[ f_k^*(\eta) \, f_k(\tau) \right] \, .
\end{equation}
Finally, the Green function $G_k(\eta, \tau)$ in Eq.~\eqref{eq: Green function} can be expressed as
\begin{equation}
    G_k(\eta, \tau) = -\frac{B_\eta^k(\tau)}{k} \theta(\eta - \tau) \, ,
    \label{eq: Green through B}
\end{equation}
by again using Eq.~\eqref{eq: Wronskian}.

To compute the induced power spectrum $\mathcal{P}_{T,\lambda}^{\mathrm{ind}}$, we follow the strategy outlined in the appendix of Ref.~\cite{garcia-bellido_2024}.
We introduce new dimensionless momentum variables,
\begin{equation}
    A = \frac{1}{2k} (p + q) \, , \qquad B = \frac{1}{2k} (p - q)
\end{equation}
for $q = |\bm{k} - \bm{p}|$. Defining
\begin{equation}
    \mathcal{I}(\tau; p, q; \alpha, \beta) = \frac{2}{\sqrt{pq}} \left[A'_\alpha(\tau, p) A'_\alpha(\tau, q) +  \alpha \beta p q A_\alpha(\tau, p) A_\alpha(\tau, q)\right] \, 
\end{equation}
and
\begin{align}
    F_\lambda(A, B; \alpha, \beta) =& \frac{1}{4}\left|\alpha \beta + 2\lambda (\alpha + \beta) A + 2\lambda (\alpha - \beta) B  \right. \\
                & \left. + 4(A^2 - B^2) + 8\lambda (\alpha - \beta) A^2 B - 8\lambda (\alpha + \beta) A B^2 - 16 \alpha \beta A^2 B^2 \right| \, \nonumber,
\end{align}
one may re-express the integral in Eq.~\eqref{eq: PT} as
\begin{align}
    \mathcal{P}_{T,\lambda}^{\mathrm{ind}}(k, \eta(N)) =& \frac{k^4}{16 \pi^4 \MP^4} \int_{A_{\mathrm{min}}}^{A_{\mathrm{max}}} \D A \int_{-B_0(A)}^{B_0(A)}  \D B \, \sum_{\alpha, \beta = \pm} F_\lambda(A, B; \alpha, \beta) \\
    &\times \left| \int_0^N \D N' \frac{k B_{\eta}^k(\tau(N'))}{e^{3N'}H(N')} \mathcal{I}\left(\tau(N'); k(A+B), k(A-B); \alpha, \beta\right)\right|^2 \,  \nonumber ,
    \label{eq: PT numerical}
\end{align}
where $N,\, N'$ count the number of $e$-folds after initialization of the background dynamics given by the GEF.
The integration boundaries of Eq.~\eqref{eq: PT numerical} are
\begin{subequations}
    \begin{align}
        A_{\mathrm{min}} =& \max\left(A_{\mathrm{IR}}, \frac{1}{2} \right), \qquad A_{\mathrm{max}} = A_{\mathrm{UV}} \\
        B_0(A)=&\left\{\begin{array}{cl}
            A-A_{\mathrm{IR}}, & \quad\text{for}\quad A_{\mathrm{min}}\leq A\leq\dfrac{1}{2}+A_{\mathrm{IR}};\\
            \dfrac{1}{2}, & \quad\text{for}\quad \dfrac{1}{2}+A_{\mathrm{IR}}<A\leq A_{\mathrm{UV}}-\dfrac{1}{2};\\
            A_{\mathrm{UV}}-A, & \quad\text{for}\quad A_{\mathrm{UV}}-\dfrac{1}{2}<A\leq A_{\mathrm{UV}}.
        \end{array}\right.  
    \end{align}
\end{subequations}
Here, $A_{\mathrm{IR}} = k_{\mathrm{IR}}/k$ and $A_{\mathrm{UV}} = k_{\mathrm{h}}(N)/k$ to ensure that we only integrate over momenta $p, q \in [k_{\mathrm{IR}}, k_{\mathrm{h}}(N)]$, with $k_{\mathrm{IR}}$ the smallest wavenumber resolved by the MbM solution. We evaluate the inner integral over the number of $e$-folds $N$ in Eq.~\eqref{eq: PT numerical} using the trapezoid rule, and the momentum integral over $A$ and $B$ by interpolation and subsequent quadrature integration.


\bibliographystyle{JHEP}
\bibliography{manuscript_v2.bib}

@article{Einstein_1916,
    author = "Einstein, Albert",
    title = "{N\"aherungsweise Integration der Feldgleichungen der Gravitation}",
    journal = "Sitzungsber. Preuss. Akad. Wiss. Berlin (Math. Phys.)",
    volume = "1916",
    pages = "688--696",
    year = "1916"
}

@article{Einstein_1918,
    author = "Einstein, Albert",
    title = {{{\"U}ber Gravitationswellen}},
    journal = "Sitzungsber. Preuss. Akad. Wiss. Berlin (Math. Phys.)",
    volume = "1918",
    pages = "154--167",
    year = "1918"
}

@article{Grishchuk_1974,
    author = "Grishchuk, L. P.",
    title = "{Amplification of gravitational waves in an istropic universe}",
    journal = "Zh. Eksp. Teor. Fiz.",
    volume = "67",
    pages = "825--838",
    year = "1974"
}

@article{Starobinsky_1979,
    author = "Starobinsky, Alexei A.",
    editor = "Khalatnikov, I. M. and Mineev, V. P.",
    title = "{Spectrum of relict gravitational radiation and the early state of the universe}",
    journal = "JETP Lett.",
    volume = "30",
    pages = "682--685",
    year = "1979"
}

@article{starobinsky_1980,
    author = "Starobinsky, Alexei A.",
    editor = "Khalatnikov, I. M. and Mineev, V. P.",
    title = "{A New Type of Isotropic Cosmological Models Without Singularity}",
    doi = "10.1016/0370-2693(80)90670-X",
    journal = "Phys. Lett. B",
    volume = "91",
    pages = "99--102",
    year = "1980"
}

@article{guth_1981,
    author = "Guth, Alan H.",
    editor = "Fang, Li-Zhi and Ruffini, R.",
    title = "{The Inflationary Universe: A Possible Solution to the Horizon and Flatness Problems}",
    reportNumber = "SLAC-PUB-2576",
    doi = "10.1103/PhysRevD.23.347",
    journal = "Phys. Rev. D",
    volume = "23",
    pages = "347--356",
    year = "1981"
}

@article{Mukhanov_1981,
    author = "Mukhanov, Viatcheslav F. and Chibisov, G. V.",
    title = "{Quantum Fluctuations and a Nonsingular Universe}",
    journal = "JETP Lett.",
    volume = "33",
    pages = "532--535",
    year = "1981"
}

@article{linde_1982,
    author = "Linde, Andrei D.",
    editor = "Fang, Li-Zhi and Ruffini, R.",
    title = "{A New Inflationary Universe Scenario: A Possible Solution of the Horizon, Flatness, Homogeneity, Isotropy and Primordial Monopole Problems}",
    reportNumber = "LEBEDEV-81-229",
    doi = "10.1016/0370-2693(82)91219-9",
    journal = "Phys. Lett. B",
    volume = "108",
    pages = "389--393",
    year = "1982"
}

@article{Mukhanov_1982,
    author = "Mukhanov, Viatcheslav F. and Chibisov, G. V.",
    title = "{The Vacuum energy and large scale structure of the universe}",
    journal = "Sov. Phys. JETP",
    volume = "56",
    pages = "258--265",
    year = "1982"
}

@article{albrecht_1982,
    author = "Albrecht, Andreas and Steinhardt, Paul J.",
    editor = "Fang, Li-Zhi and Ruffini, R.",
    title = "{Cosmology for Grand Unified Theories with Radiatively Induced Symmetry Breaking}",
    reportNumber = "UPR-0185T",
    doi = "10.1103/PhysRevLett.48.1220",
    journal = "Phys. Rev. Lett.",
    volume = "48",
    pages = "1220--1223",
    year = "1982"
}

@article{hawking_1982,
    author = "Hawking, S. W.",
    title = "{The Development of Irregularities in a Single Bubble Inflationary Universe}",
    reportNumber = "Print-83-0015 (CAMBRIDGE)",
    doi = "10.1016/0370-2693(82)90373-2",
    journal = "Phys. Lett. B",
    volume = "115",
    pages = "295",
    year = "1982"
}

@article{rubakov_1982,
    author = "Rubakov, V. A. and Sazhin, M. V. and Veryaskin, A. V.",
    title = "{Graviton Creation in the Inflationary Universe and the Grand Unification Scale}",
    doi = "10.1016/0370-2693(82)90641-4",
    journal = "Phys. Lett. B",
    volume = "115",
    pages = "189--192",
    year = "1982"
}

@article{guth_1982,
    author = "Guth, Alan H. and Pi, S. Y.",
    title = "{Fluctuations in the New Inflationary Universe}",
    doi = "10.1103/PhysRevLett.49.1110",
    journal = "Phys. Rev. Lett.",
    volume = "49",
    pages = "1110--1113",
    year = "1982"
}

@article{starobinsky_1982,
    author = "Starobinsky, Alexei A.",
    title = "{Dynamics of Phase Transition in the New Inflationary Universe Scenario and Generation of Perturbations}",
    doi = "10.1016/0370-2693(82)90541-X",
    journal = "Phys. Lett. B",
    volume = "117",
    pages = "175--178",
    year = "1982"
}

@article{fabbri_1983,
    author = "Fabbri, R. and Pollock, M. d.",
    title = "{The Effect of Primordially Produced Gravitons upon the Anisotropy of the Cosmological Microwave Background Radiation}",
    doi = "10.1016/0370-2693(83)91322-9",
    journal = "Phys. Lett. B",
    volume = "125",
    pages = "445--448",
    year = "1983"
}

@article{bardeen_1983,
    author = "Bardeen, James M. and Steinhardt, Paul J. and Turner, Michael S.",
    title = "{Spontaneous Creation of Almost Scale - Free Density Perturbations in an Inflationary Universe}",
    reportNumber = "UPR-0202T, EFI-83-13-CHICAGO",
    doi = "10.1103/PhysRevD.28.679",
    journal = "Phys. Rev. D",
    volume = "28",
    pages = "679",
    year = "1983"
}

@article{linde_1983,
    author = "Linde, Andrei D.",
    title = "{Chaotic Inflation}",
    doi = "10.1016/0370-2693(83)90837-7",
    journal = "Phys. Lett. B",
    volume = "129",
    pages = "177--181",
    year = "1983"
}

@article{abbott_1984,
    author = "Abbott, L. F. and Wise, Mark B.",
    title = "{Constraints on Generalized Inflationary Cosmologies}",
    reportNumber = "CALT-68-1100",
    doi = "10.1016/0550-3213(84)90329-8",
    journal = "Nucl. Phys. B",
    volume = "244",
    pages = "541--548",
    year = "1984"
}

@article{Salopek_1990,
    author = "Salopek, D. S. and Bond, J. R.",
    title = "{Nonlinear evolution of long wavelength metric fluctuations in inflationary models}",
    reportNumber = "FERMILAB-PUB-90-131-A",
    doi = "10.1103/PhysRevD.42.3936",
    journal = "Phys. Rev. D",
    volume = "42",
    pages = "3936--3962",
    year = "1990"
}

@article{freese_1990,
    author = "Freese, Katherine and Frieman, Joshua A. and Olinto, Angela V.",
    title = "{Natural inflation with pseudo - Nambu-Goldstone bosons}",
    reportNumber = "FERMILAB-PUB-90-177-A",
    doi = "10.1103/PhysRevLett.65.3233",
    journal = "Phys. Rev. Lett.",
    volume = "65",
    pages = "3233--3236",
    year = "1990"
}

@article{garretson_1992,
    author = "Garretson, W. Daniel and Field, George B. and Carroll, Sean M.",
    title = "{Primordial magnetic fields from pseudoGoldstone bosons}",
    eprint = "hep-ph/9209238",
    archivePrefix = "arXiv",
    reportNumber = "PRINT-92-0448 (CFA,CAMBRIDGE), CFA-3507",
    doi = "10.1103/PhysRevD.46.5346",
    journal = "Phys. Rev. D",
    volume = "46",
    pages = "5346--5351",
    year = "1992"
}

@article{liddle_1994,
    author = "Liddle, Andrew R. and Parsons, Paul and Barrow, John D.",
    title = "{Formalizing the slow roll approximation in inflation}",
    eprint = "astro-ph/9408015",
    archivePrefix = "arXiv",
    reportNumber = "SUSSEX-AST-94-8-1",
    doi = "10.1103/PhysRevD.50.7222",
    journal = "Phys. Rev. D",
    volume = "50",
    pages = "7222--7232",
    year = "1994"
}

@inproceedings{
    allen_1996,
    author = "Allen, Bruce",
    title = "{The Stochastic gravity wave background: Sources and detection}",
    booktitle = "{Les Houches School of Physics: Astrophysical Sources of Gravitational Radiation}",
    eprint = "gr-qc/9604033",
    archivePrefix = "arXiv",
    reportNumber = "WISC-MILW-96-TH-22",
    pages = "373--417",
    month = "4",
    year = "1996"
}

@article{allen_1997,
    author = "Allen, Bruce and Romano, Joseph D.",
    title = "{Detecting a stochastic background of gravitational radiation: Signal processing strategies and sensitivities}",
    eprint = "gr-qc/9710117",
    archivePrefix = "arXiv",
    reportNumber = "WISC-MILW-97-TH-14",
    doi = "10.1103/PhysRevD.59.102001",
    journal = "Phys. Rev. D",
    volume = "59",
    pages = "102001",
    year = "1999"
}

@article{maggiore_2000,
    author = "Maggiore, Michele",
    title = "{Gravitational wave experiments and early universe cosmology}",
    eprint = "gr-qc/9909001",
    archivePrefix = "arXiv",
    reportNumber = "IFUP-TH-20-99",
    doi = "10.1016/S0370-1573(99)00102-7",
    journal = "Phys. Rept.",
    volume = "331",
    pages = "283--367",
    year = "2000"
}

@article{anber_2006,
    author = "Anber, Mohamed M. and Sorbo, Lorenzo",
    title = "{N-flationary magnetic fields}",
    eprint = "astro-ph/0606534",
    archivePrefix = "arXiv",
    doi = "10.1088/1475-7516/2006/10/018",
    journal = "JCAP",
    volume = "10",
    pages = "018",
    year = "2006"
}

@article{Boyle_2008,
    author = "Boyle, Latham A. and Steinhardt, Paul J.",
    title = "{Probing the early universe with inflationary gravitational waves}",
    eprint = "astro-ph/0512014",
    archivePrefix = "arXiv",
    doi = "10.1103/PhysRevD.77.063504",
    journal = "Phys. Rev. D",
    volume = "77",
    pages = "063504",
    year = "2008"
}

@article{Anber_2010,
   title={Naturally inflating on steep potentials through electromagnetic dissipation},
   volume={81},
   ISSN={1550-2368},
   url={http://dx.doi.org/10.1103/PhysRevD.81.043534},
   DOI={10.1103/physrevd.81.043534},
   number={4},
   journal={Physical Review D},
   publisher={American Physical Society (APS)},
   author={Anber, Mohamed M. and Sorbo, Lorenzo},
   year={2010},
   month=feb 
}

@article{LIGO_2010,
	 author = "Harry, Gregory M.",
    editor = "Marka, Zsuzsa and Marka, Szabolcs",
    collaboration = "LIGO Scientific",
    title = "{Advanced LIGO: The next generation of gravitational wave detectors}",
    doi = "10.1088/0264-9381/27/8/084006",
    journal = "Class. Quant. Grav.",
    volume = "27",
    pages = "084006",
    year = "2010"
}

@article{ET_2010,
	author = "Punturo, M. and others",
    editor = "Ricci, Fulvio",
    title = "{The Einstein Telescope: A third-generation gravitational wave observatory}",
    doi = "10.1088/0264-9381/27/19/194002",
    journal = "Class. Quant. Grav.",
    volume = "27",
    pages = "194002",
    year = "2010"
}

@article{durrer_2011,
	title = {Can slow roll inflation induce relevant helical magnetic fields?},
	volume = {2011},
	issn = {1475-7516},
	url = {http://arxiv.org/abs/1005.5322},
	doi = {10.1088/1475-7516/2011/03/037},
	number = {03},
	journal = {Journal of Cosmology and Astroparticle Physics},
	author = {Durrer, Ruth and Hollenstein, Lukas and Jain, Rajeev Kumar},
	month = mar,
	year = {2011},
}

@article{Barnaby_2011_A,
    author = "Barnaby, Neil and Peloso, Marco",
    title = "{Large Nongaussianity in Axion Inflation}",
    eprint = "1011.1500",
    archivePrefix = "arXiv",
    primaryClass = "hep-ph",
    reportNumber = "UMN-TH-2926-10",
    doi = "10.1103/PhysRevLett.106.181301",
    journal = "Phys. Rev. Lett.",
    volume = "106",
    pages = "181301",
    year = "2011"
}

@article{Sorbo_2011,
    author = "Sorbo, Lorenzo",
    title = "{Parity violation in the Cosmic Microwave Background from a pseudoscalar inflaton}",
    eprint = "1101.1525",
    archivePrefix = "arXiv",
    primaryClass = "astro-ph.CO",
    doi = "10.1088/1475-7516/2011/06/003",
    journal = "JCAP",
    volume = "06",
    pages = "003",
    year = "2011"
}

@article{barnaby_2012,
    author = "Barnaby, Neil and Namba, Ryo and Peloso, Marco",
    title = "{Phenomenology of a Pseudo-Scalar Inflaton: Naturally Large Nongaussianity}",
    eprint = "1102.4333",
    archivePrefix = "arXiv",
    primaryClass = "astro-ph.CO",
    doi = "10.1088/1475-7516/2011/04/009",
    journal = "JCAP",
    volume = "04",
    pages = "009",
    year = "2011"
}

@article{cook_2012,
    author = "Cook, Jessica L. and Sorbo, Lorenzo",
    title = "{Particle production during inflation and gravitational waves detectable by ground-based interferometers}",
    eprint = "1109.0022",
    archivePrefix = "arXiv",
    primaryClass = "astro-ph.CO",
    doi = "10.1103/PhysRevD.85.023534",
    journal = "Phys. Rev. D",
    volume = "85",
    pages = "023534",
    year = "2012",
    note = "[Erratum: Phys.Rev.D 86, 069901 (2012)]"
}

@article{linde_2013,
    author = "Linde, Andrei and Mooij, Sander and Pajer, Enrico",
    title = "{Gauge field production in supergravity inflation: Local non-Gaussianity and primordial black holes}",
    eprint = "1212.1693",
    archivePrefix = "arXiv",
    primaryClass = "hep-th",
    doi = "10.1103/PhysRevD.87.103506",
    journal = "Phys. Rev. D",
    volume = "87",
    number = "10",
    pages = "103506",
    year = "2013"
}

@article{VIRGO_2014,
	author = "Acernese, F. and others",
    collaboration = "VIRGO",
    title = "{Advanced Virgo: a second-generation interferometric gravitational wave detector}",
    eprint = "1408.3978",
    archivePrefix = "arXiv",
    primaryClass = "gr-qc",
    doi = "10.1088/0264-9381/32/2/024001",
    journal = "Class. Quant. Grav.",
    volume = "32",
    number = "2",
    pages = "024001",
    year = "2015"
}

@article{LIGO_2014,
	 author = "Aasi, J. and others",
    collaboration = "LIGO Scientific",
    title = "{Advanced LIGO}",
    eprint = "1411.4547",
    archivePrefix = "arXiv",
    primaryClass = "gr-qc",
    doi = "10.1088/0264-9381/32/7/074001",
    journal = "Class. Quant. Grav.",
    volume = "32",
    pages = "074001",
    year = "2015"
}

@article{bugaev_2014,
    author = "Bugaev, Edgar and Klimai, Peter",
    title = "{Axion inflation with gauge field production and primordial black holes}",
    eprint = "1312.7435",
    archivePrefix = "arXiv",
    primaryClass = "astro-ph.CO",
    doi = "10.1103/PhysRevD.90.103501",
    journal = "Phys. Rev. D",
    volume = "90",
    number = "10",
    pages = "103501",
    year = "2014"
}

@article{ferreira_2014,
    author = "Ferreira, Ricardo Z. and Sloth, Martin S.",
    title = "{Universal Constraints on Axions from Inflation}",
    eprint = "1409.5799",
    archivePrefix = "arXiv",
    primaryClass = "hep-ph",
    reportNumber = "CP3-ORIGINS-2014-041, DIAS-2014-41",
    doi = "10.1007/JHEP12(2014)139",
    journal = "JHEP",
    volume = "12",
    pages = "139",
    year = "2014"
}

@article{bamba_2015,
    author = "Bamba, Kazuharu",
    title = "{Generation of large-scale magnetic fields, non-Gaussianity, and primordial gravitational waves in inflationary cosmology}",
    eprint = "1411.4335",
    archivePrefix = "arXiv",
    primaryClass = "astro-ph.CO",
    reportNumber = "OCHA-PP-330",
    doi = "10.1103/PhysRevD.91.043509",
    journal = "Phys. Rev. D",
    volume = "91",
    pages = "043509",
    year = "2015"
}

@article{anber_2015,
    author = "Anber, Mohamed M. and Sabancilar, Eray",
    title = "{Hypermagnetic Fields and Baryon Asymmetry from Pseudoscalar Inflation}",
    eprint = "1507.00744",
    archivePrefix = "arXiv",
    primaryClass = "hep-th",
    doi = "10.1103/PhysRevD.92.101501",
    journal = "Phys. Rev. D",
    volume = "92",
    number = "10",
    pages = "101501",
    year = "2015"
}

@article{adshead_2015,
    author = "Adshead, Peter and Giblin, John T. and Scully, Timothy R. and Sfakianakis, Evangelos I.",
    title = "{Gauge-preheating and the end of axion inflation}",
    eprint = "1502.06506",
    archivePrefix = "arXiv",
    primaryClass = "astro-ph.CO",
    doi = "10.1088/1475-7516/2015/12/034",
    journal = "JCAP",
    volume = "12",
    pages = "034",
    year = "2015"
}

@article{fujita_2015,
    author = "Fujita, Tomohiro and Namba, Ryo and Tada, Yuichiro and Takeda, Naoyuki and Tashiro, Hiroyuki",
    title = "{Consistent generation of magnetic fields in axion inflation models}",
    eprint = "1503.05802",
    archivePrefix = "arXiv",
    primaryClass = "astro-ph.CO",
    reportNumber = "IPMU-15-0029, ICRR-REPORT-699-2014-25",
    doi = "10.1088/1475-7516/2015/05/054",
    journal = "JCAP",
    volume = "05",
    pages = "054",
    year = "2015"
}

@article{abbott_2016,
author = "Abbott, B. P. and others",
    collaboration = "LIGO Scientific, Virgo",
    title = "{Observation of Gravitational Waves from a Binary Black Hole Merger}",
    eprint = "1602.03837",
    archivePrefix = "arXiv",
    primaryClass = "gr-qc",
    reportNumber = "LIGO-P150914",
    doi = "10.1103/PhysRevLett.116.061102",
    journal = "Phys. Rev. Lett.",
    volume = "116",
    number = "6",
    pages = "061102",
    year = "2016"
}

@article{Cheng_2016,
    author = "Cheng, Shu-Lin and Lee, Wolung and Ng, Kin-Wang",
    title = "{Numerical study of pseudoscalar inflation with an axion-gauge field coupling}",
    eprint = "1508.00251",
    archivePrefix = "arXiv",
    primaryClass = "astro-ph.CO",
    doi = "10.1103/PhysRevD.93.063510",
    journal = "Phys. Rev. D",
    volume = "93",
    number = "6",
    pages = "063510",
    year = "2016"
}

@article{domcke_2016,
    author = "Domcke, Valerie and Pieroni, Mauro and Bin{\'e}truy, Pierre",
    title = "{Primordial gravitational waves for universality classes of pseudoscalar inflation}",
    eprint = "1603.01287",
    archivePrefix = "arXiv",
    primaryClass = "astro-ph.CO",
    doi = "10.1088/1475-7516/2016/06/031",
    journal = "JCAP",
    volume = "06",
    pages = "031",
    year = "2016"
}

@article{adshead_2016,
    author = "Adshead, Peter and Giblin, John T. and Scully, Timothy R. and Sfakianakis, Evangelos I.",
    title = "{Magnetogenesis from axion inflation}",
    eprint = "1606.08474",
    archivePrefix = "arXiv",
    primaryClass = "astro-ph.CO",
    doi = "10.1088/1475-7516/2016/10/039",
    journal = "JCAP",
    volume = "10",
    pages = "039",
    year = "2016"
}

@article{guzzetti_2016,
    author = "Guzzetti, M. C. and Bartolo, N. and Liguori, M. and Matarrese, S.",
    title = "{Gravitational waves from inflation}",
    eprint = "1605.01615",
    archivePrefix = "arXiv",
    primaryClass = "astro-ph.CO",
    doi = "10.1393/ncr/i2016-10127-1",
    journal = "Riv. Nuovo Cim.",
    volume = "39",
    number = "9",
    pages = "399--495",
    year = "2016"
}

@article{Notari_2016,
    author = "Notari, Alessio and Tywoniuk, Konrad",
    title = "{Dissipative Axial Inflation}",
    eprint = "1608.06223",
    archivePrefix = "arXiv",
    primaryClass = "hep-th",
    reportNumber = "CERN-TH-2016-189",
    doi = "10.1088/1475-7516/2016/12/038",
    journal = "JCAP",
    volume = "12",
    pages = "038",
    year = "2016"
}

@article{LISA_2017,
    author = "Amaro-Seoane, Pau and others",
    collaboration = "LISA",
    title = "{Laser Interferometer Space Antenna}",
    eprint = "1702.00786",
    archivePrefix = "arXiv",
    primaryClass = "astro-ph.IM",
    month = "2",
    year = "2017"
}

@article{jimenez_2017,
    author = "Jim{\'e}nez, Daniel and Kamada, Kohei and Schmitz, Kai and Xu, Xun-Jie",
    title = "{Baryon asymmetry and gravitational waves from pseudoscalar inflation}",
    eprint = "1707.07943",
    archivePrefix = "arXiv",
    primaryClass = "hep-ph",
    doi = "10.1088/1475-7516/2017/12/011",
    journal = "JCAP",
    volume = "12",
    pages = "011",
    year = "2017"
}

@article{Figueroa_2018,
    author = "Figueroa, Daniel G. and Shaposhnikov, Mikhail",
    title = "{Lattice implementation of Abelian gauge theories with Chern{\textendash}Simons number and an axion field}",
    eprint = "1705.09629",
    archivePrefix = "arXiv",
    primaryClass = "hep-lat",
    reportNumber = "CERN-TH-2017-116",
    doi = "10.1016/j.nuclphysb.2017.12.001",
    journal = "Nucl. Phys. B",
    volume = "926",
    pages = "544--569",
    year = "2018"
}

@article{adshead_2018,
    author = "Adshead, Peter and Giblin, John T. and Weiner, Zachary J.",
    title = "{Gravitational waves from gauge preheating}",
    eprint = "1805.04550",
    archivePrefix = "arXiv",
    primaryClass = "astro-ph.CO",
    doi = "10.1103/PhysRevD.98.043525",
    journal = "Phys. Rev. D",
    volume = "98",
    number = "4",
    pages = "043525",
    year = "2018"
}

@article{caprini_2018,
    author = "Caprini, Chiara and Figueroa, Daniel G.",
    title = "{Cosmological Backgrounds of Gravitational Waves}",
    eprint = "1801.04268",
    archivePrefix = "arXiv",
    primaryClass = "astro-ph.CO",
    doi = "10.1088/1361-6382/aac608",
    journal = "Class. Quant. Grav.",
    volume = "35",
    number = "16",
    pages = "163001",
    year = "2018"
}

@article{domcke_2018,
    author = "Domcke, Valerie and Mukaida, Kyohei",
    title = "{Gauge Field and Fermion Production during Axion Inflation}",
    eprint = "1806.08769",
    archivePrefix = "arXiv",
    primaryClass = "hep-ph",
    reportNumber = "DESY 18-098, DESY-18-098",
    doi = "10.1088/1475-7516/2018/11/020",
    journal = "JCAP",
    volume = "11",
    pages = "020",
    year = "2018"
}

@article{
saikawa_2018,
    author = "Saikawa, Ken'ichi and Shirai, Satoshi",
    title = "{Primordial gravitational waves, precisely: The role of thermodynamics in the Standard Model}",
    eprint = "1803.01038",
    archivePrefix = "arXiv",
    primaryClass = "hep-ph",
    reportNumber = "IPMU18-0037, MPP-2018-19",
    doi = "10.1088/1475-7516/2018/05/035",
    journal = "JCAP",
    volume = "05",
    pages = "035",
    year = "2018"
}

@article{cuissa_2019,
    author = "Cuissa, Jose Roberto Canivete and Figueroa, Daniel G.",
    title = "{Lattice formulation of axion inflation. Application to preheating}",
    eprint = "1812.03132",
    archivePrefix = "arXiv",
    primaryClass = "astro-ph.CO",
    doi = "10.1088/1475-7516/2019/06/002",
    journal = "JCAP",
    volume = "06",
    pages = "002",
    year = "2019"
}

@article{CE_2019,
	 author = "Reitze, David and others",
    title = "{Cosmic Explorer: The U.S. Contribution to Gravitational-Wave Astronomy beyond LIGO}",
    eprint = "1907.04833",
    archivePrefix = "arXiv",
    primaryClass = "astro-ph.IM",
    reportNumber = "LIGO-P1900316",
    journal = "Bull. Am. Astron. Soc.",
    volume = "51",
    number = "7",
    pages = "035",
    year = "2019"
}

@misc{LISA_2019,
	author = "Baker, John and others",
    title = "{The Laser Interferometer Space Antenna: Unveiling the Millihertz Gravitational Wave Sky}",
    eprint = "1907.06482",
    archivePrefix = "arXiv",
    primaryClass = "astro-ph.IM",
    reportNumber = "FERMILAB-PUB-19-436-A",
    month = "7",
    year = "2019"
}

@article{Sobol_2019,
    author = "Sobol, O. O. and Gorbar, E. V. and Vilchinskii, S. I.",
    title = "{Backreaction of electromagnetic fields and the Schwinger effect in pseudoscalar inflation magnetogenesis}",
    eprint = "1907.10443",
    archivePrefix = "arXiv",
    primaryClass = "astro-ph.CO",
    doi = "10.1103/PhysRevD.100.063523",
    journal = "Phys. Rev. D",
    volume = "100",
    number = "6",
    pages = "063523",
    year = "2019"
}

@article{planck_2020_VI,
	author = "Aghanim, N. and others",
    collaboration = "Planck",
    title = "{Planck 2018 results. VI. Cosmological parameters}",
    eprint = "1807.06209",
    archivePrefix = "arXiv",
    primaryClass = "astro-ph.CO",
    doi = "10.1051/0004-6361/201833910",
    journal = "Astron. Astrophys.",
    volume = "641",
    pages = "A6",
    year = "2020",
    note = "[Erratum: Astron.Astrophys. 652, C4 (2021)]"
}

@article{planck_2020_X,
    author = "Akrami, Y. and others",
    collaboration = "Planck",
    title = "{Planck 2018 results. X. Constraints on inflation}",
    eprint = "1807.06211",
    archivePrefix = "arXiv",
    primaryClass = "astro-ph.CO",
    doi = "10.1051/0004-6361/201833887",
    journal = "Astron. Astrophys.",
    volume = "641",
    pages = "A10",
    year = "2020"
}

@article{adshead_2020_A,
    author = "Adshead, Peter and Giblin, John T. and Pieroni, Mauro and Weiner, Zachary J.",
    title = "{Constraining axion inflation with gravitational waves from preheating}",
    eprint = "1909.12842",
    archivePrefix = "arXiv",
    primaryClass = "astro-ph.CO",
    doi = "10.1103/PhysRevD.101.083534",
    journal = "Phys. Rev. D",
    volume = "101",
    number = "8",
    pages = "083534",
    year = "2020"
}

@article{adshead_2020_B,
    author = "Adshead, Peter and Giblin, John T. and Pieroni, Mauro and Weiner, Zachary J.",
    title = "{Constraining Axion Inflation with Gravitational Waves across 29 Decades in Frequency}",
    eprint = "1909.12843",
    archivePrefix = "arXiv",
    primaryClass = "astro-ph.CO",
    doi = "10.1103/PhysRevLett.124.171301",
    journal = "Phys. Rev. Lett.",
    volume = "124",
    number = "17",
    pages = "171301",
    year = "2020"
}

@article{Domcke_2020_Resonant,
    author = "Domcke, Valerie and Guidetti, Veronica and Welling, Yvette and Westphal, Alexander",
    title = "{Resonant backreaction in axion inflation}",
    eprint = "2002.02952",
    archivePrefix = "arXiv",
    primaryClass = "astro-ph.CO",
    reportNumber = "DESY-20-017",
    doi = "10.1088/1475-7516/2020/09/009",
    journal = "JCAP",
    volume = "09",
    pages = "009",
    year = "2020"
}

@article{Domcke_2020_Fermions,
    author = "Domcke, Valerie and Ema, Yohei and Mukaida, Kyohei",
    title = "{Chiral Anomaly, Schwinger Effect, Euler-Heisenberg Lagrangian, and application to axion inflation}",
    eprint = "1910.01205",
    archivePrefix = "arXiv",
    primaryClass = "hep-ph",
    reportNumber = "DESY-19-166, DESY 19-166",
    doi = "10.1007/JHEP02(2020)055",
    journal = "JHEP",
    volume = "02",
    pages = "055",
    year = "2020"
}

@article{saikawa_2020,
    author = "Saikawa, Ken'ichi and Shirai, Satoshi",
    title = "{Precise WIMP Dark Matter Abundance and Standard Model Thermodynamics}",
    eprint = "2005.03544",
    archivePrefix = "arXiv",
    primaryClass = "hep-ph",
    reportNumber = "IPMU20-0047, KANAZAWA-20-03, MPP-2020-62",
    doi = "10.1088/1475-7516/2020/08/011",
    journal = "JCAP",
    volume = "08",
    pages = "011",
    year = "2020"
}

@article{schmitz_2021,
    author = "Schmitz, Kai",
    title = "{New Sensitivity Curves for Gravitational-Wave Signals from Cosmological Phase Transitions}",
    eprint = "2002.04615",
    archivePrefix = "arXiv",
    primaryClass = "hep-ph",
    reportNumber = "CERN-TH-2020-018",
    doi = "10.1007/JHEP01(2021)097",
    journal = "JHEP",
    volume = "01",
    pages = "097",
    year = "2021"
}

@article{yeh_2021,
    author = "Yeh, Tsung-Han and Olive, Keith A. and Fields, Brian D.",
    title = "{The impact of new $d(p,\gamma)$3 rates on Big Bang Nucleosynthesis}",
    eprint = "2011.13874",
    archivePrefix = "arXiv",
    primaryClass = "astro-ph.CO",
    reportNumber = "UMN--TH--4004/20, FTPI--MINN--20/35",
    doi = "10.1088/1475-7516/2021/03/046",
    journal = "JCAP",
    volume = "03",
    pages = "046",
    year = "2021"
}

@article{pisanti_2021,
    author = "Pisanti, Ofelia and Mangano, Gianpiero and Miele, Gennaro and Mazzella, Pierpaolo",
    title = "{Primordial Deuterium after LUNA: concordances and error budget}",
    eprint = "2011.11537",
    archivePrefix = "arXiv",
    primaryClass = "astro-ph.CO",
    doi = "10.1088/1475-7516/2021/04/020",
    journal = "JCAP",
    volume = "04",
    pages = "020",
    year = "2021"
}

@article{LIGO_collaboration_2021,
	 author = "Abbott, R. and others",
    collaboration = "KAGRA, Virgo, LIGO Scientific",
    title = "{Upper limits on the isotropic gravitational-wave background from Advanced LIGO and Advanced Virgo{\textquoteright}s third observing run}",
    eprint = "2101.12130",
    archivePrefix = "arXiv",
    primaryClass = "gr-qc",
    reportNumber = "LIGO-DCC-P2000314",
    doi = "10.1103/PhysRevD.104.022004",
    journal = "Phys. Rev. D",
    volume = "104",
    number = "2",
    pages = "022004",
    year = "2021"
}

@article{domenech_scalar_2021,
    author = "Dom{\`e}nech, Guillem",
    title = "{Scalar Induced Gravitational Waves Review}",
    eprint = "2109.01398",
    archivePrefix = "arXiv",
    primaryClass = "gr-qc",
    doi = "10.3390/universe7110398",
    journal = "Universe",
    volume = "7",
    number = "11",
    pages = "398",
    year = "2021"
}

@article{Gorbar_2021,
    author = "Gorbar, E. V. and Schmitz, K. and Sobol, O. O. and Vilchinskii, S. I.",
    title = "{Gauge-field production during axion inflation in the gradient expansion formalism}",
    eprint = "2109.01651",
    archivePrefix = "arXiv",
    primaryClass = "hep-ph",
    reportNumber = "CERN-TH-2021-128",
    doi = "10.1103/PhysRevD.104.123504",
    journal = "Phys. Rev. D",
    volume = "104",
    number = "12",
    pages = "123504",
    year = "2021"
}

@article{bastero-gil_2022,
    author = "Bastero-Gil, Mar and Santiago, Jose and Vega-Morales, Roberto and Ubaldi, Lorenzo",
    title = "{Dark photon dark matter from a rolling inflaton}",
    eprint = "2103.12145",
    archivePrefix = "arXiv",
    primaryClass = "hep-ph",
    reportNumber = "UG-FT 333-21, CAFPE 203-21",
    doi = "10.1088/1475-7516/2022/02/015",
    journal = "JCAP",
    volume = "02",
    number = "02",
    pages = "015",
    year = "2022"
}

@article{gorbar_2022,
    author = "Gorbar, E. V. and Schmitz, K. and Sobol, O. O. and Vilchinskii, S. I.",
    title = "{Hypermagnetogenesis from axion inflation: Model-independent estimates}",
    eprint = "2111.04712",
    archivePrefix = "arXiv",
    primaryClass = "hep-ph",
    reportNumber = "CERN-TH-2021-185",
    doi = "10.1103/PhysRevD.105.043530",
    journal = "Phys. Rev. D",
    volume = "105",
    number = "4",
    pages = "043530",
    year = "2022"
}

@article{abazajian_cmb-s4_2022,
	 author = "Abazajian, Kevork and others",
    collaboration = "CMB-S4",
    title = "{CMB-S4: Forecasting Constraints on Primordial Gravitational Waves}",
    eprint = "2008.12619",
    archivePrefix = "arXiv",
    primaryClass = "astro-ph.CO",
    reportNumber = "FERMILAB-PUB-20-468-AE-SCD",
    doi = "10.3847/1538-4357/ac1596",
    journal = "Astrophys. J.",
    volume = "926",
    number = "1",
    pages = "54",
    year = "2022"
}

@article{tristram_2022,
	author = "Tristram, M. and others",
    title = "{Improved limits on the tensor-to-scalar ratio using BICEP and Planck data}",
    eprint = "2112.07961",
    archivePrefix = "arXiv",
    primaryClass = "astro-ph.CO",
    doi = "10.1103/PhysRevD.105.083524",
    journal = "Phys. Rev. D",
    volume = "105",
    number = "8",
    pages = "083524",
    year = "2022"
}

@article{caravano_2022,
    author = "Caravano, Angelo and Komatsu, Eiichiro and Lozanov, Kaloian D. and Weller, Jochen",
    title = "{Lattice simulations of Abelian gauge fields coupled to axions during inflation}",
    eprint = "2110.10695",
    archivePrefix = "arXiv",
    primaryClass = "astro-ph.CO",
    doi = "10.1103/PhysRevD.105.123530",
    journal = "Phys. Rev. D",
    volume = "105",
    number = "12",
    pages = "123530",
    year = "2022"
}

@article{fujita_2022,
    author = "Fujita, Tomohiro and Kume, Jun'ya and Mukaida, Kyohei and Tada, Yuichiro",
    title = "{Effective treatment of U(1) gauge field and charged particles in axion inflation}",
    eprint = "2204.01180",
    archivePrefix = "arXiv",
    primaryClass = "hep-ph",
    reportNumber = "RESCEU-3/22, KEK-TH-2402",
    doi = "10.1088/1475-7516/2022/09/023",
    journal = "JCAP",
    volume = "09",
    pages = "023",
    year = "2022"
}

@article{cado_2022,
    author = "Cado, Yann and Quir{\'o}s, Mariano",
    title = "{Numerical study of the Schwinger effect in axion inflation}",
    eprint = "2208.10977",
    archivePrefix = "arXiv",
    primaryClass = "hep-ph",
    doi = "10.1103/PhysRevD.106.123527",
    journal = "Phys. Rev. D",
    volume = "106",
    number = "12",
    pages = "123527",
    year = "2022"
}

@article{domcke_2023,
    author = "Domcke, Valerie and Kamada, Kohei and Mukaida, Kyohei and Schmitz, Kai and Yamada, Masaki",
    title = "{Wash-in leptogenesis after axion inflation}",
    eprint = "2210.06412",
    archivePrefix = "arXiv",
    primaryClass = "hep-ph",
    reportNumber = "CERN-TH-2022-162, RESCEU-17/22, KEK-TH-2455, MS-TP-22-37, TU-1170",
    doi = "10.1007/JHEP01(2023)053",
    journal = "JHEP",
    volume = "01",
    pages = "053",
    year = "2023"
}

@article{ozsoy_2023,
    author = {{\"O}zsoy, Ogan and Tasinato, Gianmassimo},
    title = "{Inflation and Primordial Black Holes}",
    eprint = "2301.03600",
    archivePrefix = "arXiv",
    primaryClass = "astro-ph.CO",
    doi = "10.3390/universe9050203",
    journal = "Universe",
    volume = "9",
    number = "5",
    pages = "203",
    year = "2023"
}

@article{gorbar_2023,
    author = "Gorbar, E. V. and Momot, A. I. and Rudenok, I. V. and Sobol, O. O. and Vilchinskii, S. I. and Oleinikova, I. V.",
    title = "{Chirality Production during Axion Inflation}",
    eprint = "2111.05848",
    archivePrefix = "arXiv",
    primaryClass = "hep-ph",
    doi = "10.15407/ujpe68.11.717",
    journal = "Ukr. J. Phys.",
    volume = "68",
    number = "11",
    pages = "717",
    year = "2023"
}

@article{reardon_PPTA_2023,
	 author = "Reardon, Daniel J. and others",
    title = "{Search for an Isotropic Gravitational-wave Background with the Parkes Pulsar Timing Array}",
    eprint = "2306.16215",
    archivePrefix = "arXiv",
    primaryClass = "astro-ph.HE",
    doi = "10.3847/2041-8213/acdd02",
    journal = "Astrophys. J. Lett.",
    volume = "951",
    number = "1",
    pages = "L6",
    year = "2023"
}

@article{xu_CPTA_2023,
	author = "Xu, Heng and others",
    title = "{Searching for the Nano-Hertz Stochastic Gravitational Wave Background with the Chinese Pulsar Timing Array Data Release I}",
    eprint = "2306.16216",
    archivePrefix = "arXiv",
    primaryClass = "astro-ph.HE",
    doi = "10.1088/1674-4527/acdfa5",
    journal = "Res. Astron. Astrophys.",
    volume = "23",
    number = "7",
    pages = "075024",
    year = "2023"
}

@article{agazie_nanograv_2023,
	author = "Agazie, Gabriella and others",
    collaboration = "NANOGrav",
    title = "{The NANOGrav 15 yr Data Set: Evidence for a Gravitational-wave Background}",
    eprint = "2306.16213",
    archivePrefix = "arXiv",
    primaryClass = "astro-ph.HE",
    doi = "10.3847/2041-8213/acdac6",
    journal = "Astrophys. J. Lett.",
    volume = "951",
    number = "1",
    pages = "L8",
    year = "2023"
}

@article{afzal_nanograv_NP_2023,
	 author = "Afzal, Adeela and others",
    collaboration = "NANOGrav",
    title = "{The NANOGrav 15 yr Data Set: Search for Signals from New Physics}",
    eprint = "2306.16219",
    archivePrefix = "arXiv",
    primaryClass = "astro-ph.HE",
    reportNumber = "FERMILAB-PUB-23-589-T",
    doi = "10.3847/2041-8213/acdc91",
    journal = "Astrophys. J. Lett.",
    volume = "951",
    number = "1",
    pages = "L11",
    year = "2023",
    note = "[Erratum: Astrophys. J. Lett. 971, L27 (2024)]"
}

@article{epta+inpta_2023,
	author = "Antoniadis, J. and others",
    collaboration = "EPTA, InPTA:",
    title = "{The second data release from the European Pulsar Timing Array - III. Search for gravitational wave signals}",
    eprint = "2306.16214",
    archivePrefix = "arXiv",
    primaryClass = "astro-ph.HE",
    doi = "10.1051/0004-6361/202346844",
    journal = "Astron. Astrophys.",
    volume = "678",
    pages = "A50",
    year = "2023"
}

@article{peloso_2023,
    author = "Peloso, Marco and Sorbo, Lorenzo",
    title = "{Instability in axion inflation with strong backreaction from gauge modes}",
    eprint = "2209.08131",
    archivePrefix = "arXiv",
    primaryClass = "astro-ph.CO",
    doi = "10.1088/1475-7516/2023/01/038",
    journal = "JCAP",
    volume = "01",
    pages = "038",
    year = "2023"
}

@article{Figueroa_2023,
    author = "Figueroa, Daniel G. and Lizarraga, Joanes and Urio, Ander and Urrestilla, Jon",
    title = "{Strong Backreaction Regime in Axion Inflation}",
    eprint = "2303.17436",
    archivePrefix = "arXiv",
    primaryClass = "astro-ph.CO",
    doi = "10.1103/PhysRevLett.131.151003",
    journal = "Phys. Rev. Lett.",
    volume = "131",
    number = "15",
    pages = "151003",
    year = "2023"
}

@article{caravano_2023,
    author = "Caravano, Angelo and Komatsu, Eiichiro and Lozanov, Kaloian D. and Weller, Jochen",
    title = "{Lattice simulations of axion-U(1) inflation}",
    eprint = "2204.12874",
    archivePrefix = "arXiv",
    primaryClass = "astro-ph.CO",
    doi = "10.1103/PhysRevD.108.043504",
    journal = "Phys. Rev. D",
    volume = "108",
    number = "4",
    pages = "043504",
    year = "2023"
}

@article{durrer_2023,
    author = "Durrer, R. and Sobol, O. and Vilchinskii, S.",
    title = "{Backreaction from gauge fields produced during inflation}",
    eprint = "2303.04583",
    archivePrefix = "arXiv",
    primaryClass = "gr-qc",
    reportNumber = "MS-TP-23-06",
    doi = "10.1103/PhysRevD.108.043540",
    journal = "Phys. Rev. D",
    volume = "108",
    number = "4",
    pages = "043540",
    year = "2023"
}

@article{garcia-bellido_2024,
    author = "Garc\'{i}a-Bellido, Juan and Papageorgiou, Alexandros and Peloso, Marco and Sorbo, Lorenzo",
    title = "{A flashing beacon in axion inflation: recurring bursts of gravitational waves in the strong backreaction regime}",
    eprint = "2303.13425",
    archivePrefix = "arXiv",
    primaryClass = "astro-ph.CO",
    doi = "10.1088/1475-7516/2024/01/034",
    journal = "JCAP",
    volume = "01",
    pages = "034",
    year = "2024"
}

@article{eckardstein_2023,
    author = "von Eckardstein, Richard and Peloso, Marco and Schmitz, Kai and Sobol, Oleksandr and Sorbo, Lorenzo",
    title = "{Axion inflation in the strong-backreaction regime: decay of the Anber-Sorbo solution}",
    eprint = "2309.04254",
    archivePrefix = "arXiv",
    primaryClass = "hep-ph",
    reportNumber = "ACFI-T23-05, MS-TP-23-38",
    doi = "10.1007/JHEP11(2023)183",
    journal = "JHEP",
    volume = "11",
    pages = "183",
    year = "2023"
}

@article{domcke_2024,
    author = "Domcke, Valerie and Ema, Yohei and Sandner, Stefan",
    title = "{Perturbatively including inhomogeneities in axion inflation}",
    eprint = "2310.09186",
    archivePrefix = "arXiv",
    primaryClass = "astro-ph.CO",
    reportNumber = "IFIC/23-45, FTUV-23-1005.0503, UMN-TH-4226/23, FTPI-MINN-23-18, CERN-TH-2023-186",
    doi = "10.1088/1475-7516/2024/03/019",
    journal = "JCAP",
    volume = "03",
    pages = "019",
    year = "2024"
}

@article{drewes_2024,
    author = "Drewes, Marco and Georis, Yannis and Klasen, Michael and Wiggering, Luca Paolo and Wong, Yvonne Y. Y.",
    title = "{Towards a precision calculation of N $_{eff}$ in the Standard Model. Part III. Improved estimate of NLO contributions to the collision integral}",
    eprint = "2402.18481",
    archivePrefix = "arXiv",
    primaryClass = "hep-ph",
    reportNumber = "CPPC-2024-01, MS-TP-24-06",
    doi = "10.1088/1475-7516/2024/06/032",
    journal = "JCAP",
    volume = "06",
    pages = "032",
    year = "2024"
}

@article{corba_2024,
    author = "Corb{\`a}, Sofia P. and Sorbo, Lorenzo",
    title = "{Correlated scalar perturbations and gravitational waves from axion inflation}",
    eprint = "2403.03338",
    archivePrefix = "arXiv",
    primaryClass = "astro-ph.CO",
    doi = "10.1088/1475-7516/2024/10/024",
    journal = "JCAP",
    volume = "10",
    pages = "024",
    year = "2024"
}

@article{durrer_2024,
    author = "Durrer, R. and von Eckardstein, R. and Garg, Deepen and Schmitz, K. and Sobol, O. and Vilchinskii, S.",
    title = "{Scalar perturbations from inflation in the presence of gauge fields}",
    eprint = "2404.19694",
    archivePrefix = "arXiv",
    primaryClass = "astro-ph.CO",
    reportNumber = "MS-TP-24-10",
    doi = "10.1103/PhysRevD.110.043533",
    journal = "Phys. Rev. D",
    volume = "110",
    number = "4",
    pages = "043533",
    year = "2024"
}

@article{bastero-gil_2024_A,
    author = "Bastero-Gil, Mar and Ferraz, Paulo B. and Ubaldi, Lorenzo and Vega-Morales, Roberto",
    title = "{Schwinger dark matter production}",
    eprint = "2312.15137",
    archivePrefix = "arXiv",
    primaryClass = "hep-ph",
    reportNumber = "UG-FT 329-23, CAFPE 199-23, SISSA 44/2020/FISI, CA21106",
    doi = "10.1088/1475-7516/2024/10/078",
    journal = "JCAP",
    volume = "10",
    pages = "078",
    year = "2024"
}

@article{bastero-gil_2024_B,
    author = "Bastero-Gil, Mar and Ferraz, Paulo B. and Ubaldi, Lorenzo and Vega-Morales, Roberto",
    title = "{Super heavy dark matter from inflationary Schwinger production}",
    eprint = "2311.09475",
    archivePrefix = "arXiv",
    primaryClass = "hep-ph",
    reportNumber = "UG-FT 328-23, CAFPE 198-23, CA21106",
    doi = "10.1103/PhysRevD.110.095019",
    journal = "Phys. Rev. D",
    volume = "110",
    number = "9",
    pages = "095019",
    year = "2024"
}

@article{galloni_2024,
    author = "Galloni, Giacomo and Henrot-Versill{\'e}, Sophie and Tristram, Matthieu",
    title = "{Robust constraints on tensor perturbations from cosmological data: A comparative analysis from Bayesian and frequentist perspectives}",
    eprint = "2405.04455",
    archivePrefix = "arXiv",
    primaryClass = "astro-ph.CO",
    doi = "10.1103/PhysRevD.110.063511",
    journal = "Phys. Rev. D",
    volume = "110",
    number = "6",
    pages = "063511",
    year = "2024"
}

@article{Figueroa_2024,
    author = "Figueroa, Daniel G. and Lizarraga, Joanes and Loayza, Nicol{\'a}s and Urio, Ander and Urrestilla, Jon",
    title = "{Nonlinear dynamics of axion inflation: A detailed lattice study}",
    eprint = "2411.16368",
    archivePrefix = "arXiv",
    primaryClass = "astro-ph.CO",
    doi = "10.1103/PhysRevD.111.063545",
    journal = "Phys. Rev. D",
    volume = "111",
    number = "6",
    pages = "063545",
    year = "2025"
}

@article{greco_2024,
    author = "Greco, Federico and Peloso, Marco",
    title = "{Analytic results in aligned axion inflation}",
    eprint = "2409.01126",
    archivePrefix = "arXiv",
    primaryClass = "astro-ph.CO",
    doi = "10.1088/1475-7516/2025/01/074",
    journal = "JCAP",
    volume = "01",
    pages = "074",
    year = "2025"
}

@article{miles_meerkat_2024,
	author = "Miles, Matthew T. and others",
    title = "{The MeerKAT Pulsar Timing Array: the first search for gravitational waves with the MeerKAT radio telescope}",
    eprint = "2412.01153",
    archivePrefix = "arXiv",
    primaryClass = "astro-ph.HE",
    doi = "10.1093/mnras/stae2571",
    journal = "Mon. Not. Roy. Astron. Soc.",
    volume = "536",
    number = "2",
    pages = "1489--1500",
    year = "2024"
}

@article{agazie_nanograv_RPL_2025,
	 author = "Agazie, Gabriella and others",
    title = "{The NANOGrav 15 yr Data Set: Running of the Spectral Index}",
    eprint = "2408.10166",
    archivePrefix = "arXiv",
    primaryClass = "astro-ph.HE",
    doi = "10.3847/2041-8213/ad99d3",
    journal = "Astrophys. J. Lett.",
    volume = "978",
    number = "2",
    pages = "L29",
    year = "2025"
}

@article{eckardstein_2025,
    author = "von Eckardstein, Richard and Schmitz, Kai and Sobol, Oleksandr",
    title = "{On the Schwinger effect during axion inflation}",
    eprint = "2408.16538",
    archivePrefix = "arXiv",
    primaryClass = "hep-ph",
    reportNumber = "MS-TP-24-20",
    doi = "10.1007/JHEP02(2025)096",
    journal = "JHEP",
    volume = "02",
    pages = "096",
    year = "2025"
}

@article{sharma_2025,
    author = "Sharma, Ramkishor and Brandenburg, Axel and Subramanian, Kandaswamy and Vikman, Alexander",
    title = "{Lattice simulations of axion-U(1) inflation: gravitational waves, magnetic fields, and scalar statistics}",
    eprint = "2411.04854",
    archivePrefix = "arXiv",
    primaryClass = "astro-ph.CO",
    reportNumber = "NORDITA-2024-040",
    doi = "10.1088/1475-7516/2025/05/079",
    journal = "JCAP",
    volume = "05",
    pages = "079",
    year = "2025"
}

@article{Talebian_2025,
    author = "Talebian, Alireza and Firouzjahi, Hassan",
    title = "{Axion USR Inflation}",
    eprint = "2507.02685",
    archivePrefix = "arXiv",
    primaryClass = "astro-ph.CO",
    month = "7",
    year = "2025"
}

@article{iarygina_2025,
    author = "Iarygina, Oksana and Sfakianakis, Evangelos I. and Brandenburg, Axel",
    title = "{Schwinger effect in axion inflation on a lattice}",
    eprint = "2506.20538",
    archivePrefix = "arXiv",
    primaryClass = "astro-ph.CO",
    reportNumber = "NORDITA-2025-030",
    month = "6",
    year = "2025"
}

@article{Planck:2015zfm,
    author = "Ade, P. A. R. and others",
    collaboration = "Planck",
    title = "{Planck 2015 results. XVII. Constraints on primordial non-Gaussianity}",
    eprint = "1502.01592",
    archivePrefix = "arXiv",
    primaryClass = "astro-ph.CO",
    doi = "10.1051/0004-6361/201525836",
    journal = "Astron. Astrophys.",
    volume = "594",
    pages = "A17",
    year = "2016"
}

@article{Planck:2019kim,
    author = "Akrami, Y. and others",
    collaboration = "Planck",
    title = "{Planck 2018 results. IX. Constraints on primordial non-Gaussianity}",
    eprint = "1905.05697",
    archivePrefix = "arXiv",
    primaryClass = "astro-ph.CO",
    doi = "10.1051/0004-6361/201935891",
    journal = "Astron. Astrophys.",
    volume = "641",
    pages = "A9",
    year = "2020"
}

@book{dodelsonModernCosmology2021,
  title = {{Modern Cosmology}},
  author = {Dodelson, Scott and Schmidt, Fabian},
  year = {2021},
  edition = {Second edition},
  publisher = {Academic Press},
  address = {London, United Kingdom},
  isbn = {978-0-12-815948-4}
}

@article{Jamieson:2025ngu,
    author = "Jamieson, Drew and Caravano, Angelo and Komatsu, Eiichiro",
    title = "{Primordial Power Spectrum and Bispectrum from Lattice Simulations of Axion-U(1) Inflation}",
    eprint = "2507.22285",
    archivePrefix = "arXiv",
    primaryClass = "astro-ph.CO",
    month = "7",
    year = "2025"
}

@article{eckardstein_2025a,
    author = "von Eckardstein, Richard and Schmitz, Kai and Sobol, Oleksandr",
    title = "{Gravitational waves from axion inflation in the gradient expansion formalism. Part II. Fermionic axion inflation}",
    eprint = "2509.25013",
    archivePrefix = "arXiv",
    primaryClass = "astro-ph.CO",
    reportNumber = "MS-TP-25-27",
    doi = "10.1007/JHEP03(2026)072",
    journal = "JHEP",
    volume = "03",
    pages = "072",
    year = "2026"
}

@article{garcia-bellido_2016,
    author = "Garc\'{i}a-Bellido, Juan and Peloso, Marco and {\"U}nal, Caner",
    title = "{Gravitational waves at interferometer scales and primordial black holes in axion inflation}",
    eprint = "1610.03763",
    archivePrefix = "arXiv",
    primaryClass = "astro-ph.CO",
    reportNumber = "IFT-UAM-CSIC-16-100, UMN-TH-3607-16",
    doi = "10.1088/1475-7516/2016/12/031",
    journal = "JCAP",
    volume = "12",
    pages = "031",
    year = "2016"
}

@article{ozsoy_2020,
    author = {{\"O}zsoy, Ogan},
    title = "{Synthetic Gravitational Waves from a Rolling Axion Monodromy}",
    eprint = "2005.10280",
    archivePrefix = "arXiv",
    primaryClass = "astro-ph.CO",
    doi = "10.1088/1475-7516/2021/04/040",
    journal = "JCAP",
    volume = "04",
    pages = "040",
    year = "2021"
}

@article{unal_2023,
    author = "{\"U}nal, Caner and Papageorgiou, Alexandros and Obata, Ippei",
    title = "{Axion-gauge dynamics during inflation as the origin of pulsar timing array signals and primordial black holes}",
    eprint = "2307.02322",
    archivePrefix = "arXiv",
    primaryClass = "astro-ph.CO",
    doi = "10.1016/j.physletb.2024.138873",
    journal = "Phys. Lett. B",
    volume = "856",
    pages = "138873",
    year = "2024"
}

@article{vonEckardstein:2025jug,
        author        = "von Eckardstein, Richard",
        title         = "GEFF: The Gradient Expansion Formalism Factory -- A tool for inflationary gauge-field production",
        eprint        = "2510.12644",
        archivePrefix = "arXiv",
        primaryClass  = "astro-ph.CO",
        year          = "2025"}




\end{document}